\documentclass[jcp, twocolumn, superscriptaddress]{revtex4-2}
\newcommand{\SingleColFigScale}{1.0} 
\newcommand{\DoubleColFigScale}{1.0} 

\usepackage{graphicx}
\usepackage[usenames,dvipsnames]{xcolor}
\usepackage{amsmath}
\usepackage{amssymb}
\usepackage{mathtools}
\usepackage{accents}
\usepackage{bm}
\usepackage{multirow}
\usepackage[pdfusetitle]{hyperref}
\definecolor{link}{rgb}{0.07, 0.07, 0.80}
\hypersetup{
colorlinks=true,
linkcolor=link,
citecolor=link,
urlcolor=link
}

\makeatletter
\def\blfootnote{\gdef\@thefnmark{}\@footnotetext}
\makeatother

\begin{document}

\title{
Collective bath coordinate mapping of ``hierarchy'' in hierarchical equations of motion
}


\author{Tatsushi Ikeda}
\email{t\_ikeda@chemsys.t.u-tokyo.ac.jp}
\affiliation{Department of Chemical System Engineering, The University of Tokyo, Tokyo 113-8656, Japan}

\author{Akira Nakayama}
\affiliation{Department of Chemical System Engineering, The University of Tokyo, Tokyo 113-8656, Japan}

\date{\today}

\begin{abstract}
  The theory of hierarchical equations of motion (HEOM) is one of the standard methods to give exact evaluations of the dynamics as coupled to harmonic oscillator environments.
  However, the theory is numerically demanding due to its hierarchy, which is the set of auxiliary elements introduced to capture the non-Markovian and non-perturbative effects of environments.
  When system-bath coupling becomes relatively strong, the required computational resources and precision move beyond the regime that can be currently handled.
  This article presents a new representation of HEOM theory in which the hierarchy is mapped into a continuous space of a collective bath coordinate and several auxiliary coordinates as the form of the quantum Fokker--Planck equation.
  This representation gives a rigorous time evolution of the bath coordinate distribution and is more stable and efficient than the original HEOM theory, particularly when there is a strong system-bath coupling.
  We demonstrate the suitability of this approach to treat vibronic system models coupled to environments.

  \noindent\hrulefill
  
  This article may be downloaded for personal use only.
  Any other use requires prior permission of the authors and AIP Publishing.
  This article appeared in [T.~Ikeda and A.~Nakayama, J.~Chem.~Phys.~\textbf{156}, 104104 (2022)] and may be found at \href{https://doi.org/10.1063/5.0082936}{https://doi.org/10.1063/5.0082936}.
\end{abstract}

\pacs{Valid PACS appear here}


\maketitle

\section{INTRODUCTION}
\label{sec:introduction}

Theories that describe open quantum dynamics are widely utilized to understand chemical and biological phenomena as energy transfer between systems and environments has an essential role in their behaviors.\cite{may2008book, weiss2011book}
The descriptions of ultra-fast phenomena, such as those observed in recent advanced spectroscopy experiments, \cite{kreisbeck2012jpcl, prokhorenko2016jpcl, miyata2017mc, scholes2017nature, scholes2018jpcl, rafiq2021nc} sometimes require employing open quantum theories that can address the non-Markovian and non-perturbative effects of environments.
When the system dynamics are relatively fast, the timescale could be similar to the correlation timescales of the surrounding environment, which is not suitable for the Markovian approximation.
Many theories have been developed to capture the system-environment coupling effects, \cite{tanimura1994jcp, makri1992cpl, ilk1994jcp, wang2000jcp, wang2007jpca, chin2010jmp, prior2010prl, xie2014jcp} and the theory of hierarchical equations of motion (HEOM) is one of the standard approaches.\cite{tanimura1989jpsj, tanimura2006jpsj, xu2005jcp, tang2015jcp, ikeda2020jcp, tanimura2020jcp}
The HEOM theory is used not only to describe quantum dynamics\cite{ishizaki2008cp, tanimura2009acr, kato2016jcp, chen2019jpcl, mangaud2019jcp, takahashi2020jpsj, zhangj2020jcp, iwamoto2021jce, xu2021prb} but also as reference calculations for theoretical developments.\cite{subotnik2016arpc, lambert2019nc}

The HEOM theory was originally developed by Tanimura and Kubo to describe the dynamics of an open quantum system that is coupled to a high-temperature Drude bosonic environment (bath), which could be characterized with a single-exponential environment correlation function.\cite{tanimura1989jpsj}
Several extensions have been performed to treat more general situations.
For example, treatments for low-temperature quantum effects come from the Bose--Einstein distribution function,\cite{ishizaki2005jpsj, hu2011jcp, zhang2020jcp} multi-exponential bath correlation functions,\cite{tanimura1990pra, tanimura1994jpsj, tanaka2009jpsj, tanimura2012jcp, fujihashi2015jcp, dunn2019jcp} and non-exponential correlation functions.\cite{xu2005jcp, tang2015jcp, ikeda2020jcp}
However, these theories are numerically demanding due to their hierarchical structures, which is a set of auxiliary elements introduced to capture the non-Markovian and non-perturbative effects of environments.
A stronger system-bath coupling in the model system creates a greater hierarchy to obtain converged calculations.
When the coupling becomes relatively strong, the required computational resources and precision grow beyond the regime we can currently handle.
As seen in the numerical examples given in this paper, coupling easily reaches this condition, especially in the case of vibronic systems.

This article presents a representation of HEOM theory in which the hierarchy is mapped into a continuous space of collective bath coordinates and several auxiliary coordinates.
This approach is more stable and efficient than the original HEOM theory, particularly when strong system-bath coupling exists.
It has been previously shown that HEOM for high-temperature Drude bosonic baths is equivalent to a modified version of the Zusman equation,\cite{tanimura1989jpsj, garg1985jcp, shi2009jcp2, ikeda2019jctc} while HEOM for undamped bath modes is equivalent to the Wigner transport equation as coupled to the system.\cite{kapral1999jcp, liu2014jcp, yan2020jcp}
This paper shows the treatment for damped bath correlation functions, including quantum low-temperature effects.
Our exact derivation gives a Smoluchowski-type quantum Fokker--Planck equation for auxiliary modes as coupled to the system.\cite{ikeda2019jctc}
A concept that uses similar auxiliary modes is shown in Refs.~\onlinecite{tamascelli2018prl, lambert2019nc} but they assumed a Lindblad form.

The organization of this paper is given as follows.
In section \ref{sec:theory}, we map the hierarchy in the HEOM into a continuous coordinate space and construct the collective bath coordinate representation for a Drude spectral density model.
In section \ref{sec:results}, we present the numerical results for the cases of a vibronic system and a two-level system, which shows the pros and cons of our proposed theory.
Finally, section \ref{sec:conclusion} is devoted to concluding remarks.

\section{THEORY}
\label{sec:theory}

While it is possible to construct our approach without using HEOM theory, we derive our approach from HEOM theory as this derivation clarifies the relationship between the two theories and gives another perspective of the physical meanings of the hierarchical elements in HEOM theory.

\subsection{Hamiltonian}

We consider a system that is linearly coupled to a harmonic oscillator bath.
We assume that the system-bath interactions are characterized from a single system subspace operator $\hat{V}$ for simplicity.
The total Hamiltonian of the system is expressed as
\begin{align}
  \hat{H}^{\mathrm{tot}}&\equiv \hat{H}+\hat{H}^{\mathrm{bath}}+\hat{H}^{\mathrm{int}}
  \label{eq:total-hamiltonian},
\end{align}
where $\hat{H}$, $\hat{H}^{\mathrm{bath}}$, and $\hat{H}^{\mathrm{int}}$ are the Hamiltonian operators for the system, bath, and interactions, respectively.
\begin{subequations}
  The bath Hamiltonian is given as
  \begin{align}
    \hat{H}^{\mathrm{bath}}\equiv \sum _{\xi }\frac{\hbar \omega _{\xi }}{2}\left(\hat{p}_{\xi }^{2}+\hat{x}_{\xi }^{2}\right),
  \end{align}
  where $\hat{x}_{\xi }$, $\hat{p}_{\xi }$, and $\omega _{\xi }$ are the dimensionless coordinate, conjugate momentum, and characteristic frequency of the $\xi $th bath mode, respectively. The interaction Hamiltonian is expressed as
  \begin{align}
    \hat{H}^{\mathrm{int}}\equiv -\hbar \hat{V}\sum _{\xi }g_{\xi }\hat{x}_{\xi }=-\hbar \hat{V}\hat{X}
    \label{eq:H-int}
  \end{align}
  where $g_{\xi }$ is the coupling strength between the system and the $\xi $th bath mode.
  We introduce the collective bath coordinate as $\hat{X}\equiv \sum _{\xi }g_{\xi }\hat{x}_{\xi }$.
  Note that the bath modes are only coupled to the system through the coordinate $\hat{X}$.
\end{subequations}

\subsection{Bath}
We assume that the total density operator at $t=t_{0}$ can be written as
\begin{align}
  \hat{\rho }^{\mathrm{tot}}(t_{0})=\hat{\rho }(t_{0}){\otimes }\hat{\rho }_{\mathrm{eq}}^{\mathrm{bath}},
  \label{eq:fact-init}
\end{align}
where $\hat{\rho }(t_{0})$ is the reduced density operator of the system subspace and $\hat{\rho }_{\mathrm{eq}}^{\mathrm{bath}}$ is the bath equilibrium density operator at the temperature $T$, i.e., $\hat{\rho }_{\mathrm{eq}}^{\mathrm{bath}}=e^{-\beta \hat{H}^{\mathrm{bath}}}/\mathcal{Z}^{\mathrm{bath}}$.
Here, $\mathcal{Z}^{\mathrm{bath}}$ is the partition function of the bath and $\beta \equiv 1/k_{\mathrm{B}}T$ is the inverse temperature divided by Boltzmann's constant $k_{\mathrm{B}}$.
Open quantum theories, including HEOM theory, are concerned primarily with the dynamic properties of the reduced density operator at $t>t_{0}$, which is defined as
\begin{align}
  \hat{\rho }(t)\equiv \mathrm{Tr}^\mathrm{bath}\{\hat{\rho }^{\mathrm{tot}}(t)\},
\end{align}
where $\mathrm{Tr}^\mathrm{bath}\{\dots \}$ is the trace over the bath subspace.

Note that the factorized initial condition of Eq.~\eqref{eq:fact-init} is temporarily introduced to evaluate the time evolution of the reduced system and is not a restriction of the numerical calculations.
If we need to start from the correlated thermal equilibrium state of the total system, one of the possible prescriptions is to simulate the time evolution of the system from $t=t_{0}$ with the temporal initial state of Eq.~\eqref{eq:fact-init} to a sufficiently long time $t=t_{\mathrm{i}}\gg t_{0}$ at which the total system reaches the thermal equilibrium state.
We can then regard the state at $t=t_{\mathrm{i}}$ as the initial state for the following calculations.\cite{tanimura2006jpsj, tanimura2012jcp}
For other possible prescriptions, see Refs.~\onlinecite{tanimura2014jcp, tanimura2015jcp, zhang2017jcp}.

For a harmonic environment model, all the effects of the set of bath modes to the system dynamics are characterized entirely by its spectral density, which is defined as
\begin{align}
  \mathcal{J}(\omega )\equiv \pi \sum _{\xi }\frac{g_{\xi }^{2}}{2}\delta \left(\omega -\omega _{\xi }\right),
\end{align}
where $\delta (\omega )$ is the Dirac delta function.
The symmetrized correlation function $\mathcal{S}(t)$ and anti-symmetrized correlation function $\mathcal{A}(t)$ of the collective bath coordinate $\hat{X}$ are expressed as
\begin{subequations}
  \begin{align}
    \mathcal{S}(t)&\equiv \frac{1}{2}\left\langle \left\{\hat{X}(t),\hat{X}(0)\right\}\right\rangle ^{\mathrm{bath}}\notag\\
    &=\frac{2}{\pi }\int _{0}^{\infty }\!d\omega \,\mathcal{J}(\omega )\left(n^{\mathrm{BE}}(\omega ,T)+\frac{1}{2}\right)\cos \omega t
    \label{eq:S}\\
    \intertext{and}
    \mathcal{A}(t)&\equiv \frac{1}{2i}\left\langle \left[\hat{X}(t),\hat{X}(0)\right]\right\rangle ^{\mathrm{bath}}\notag\\
    &=-\frac{1}{\pi }\int _{0}^{\infty }\!d\omega \,\mathcal{J}(\omega )\sin \omega t
    \label{eq:A}.
  \end{align}
\end{subequations}
Here, $[\hat{A},\hat{B}]\equiv \hat{A}\hat{B}-\hat{B}\hat{A}$, $\{\hat{A},\hat{B}\}\equiv \hat{A}\hat{B}+\hat{B}\hat{A}$, $\langle \dots \rangle ^{\mathrm{bath}}\equiv \mathrm{Tr}^\mathrm{bath}\{\dots \hat{\rho }_{\mathrm{eq}}^{\mathrm{bath}}\}$, and $n^{\mathrm{BE}}(\omega ,T)=(e^{\beta \hbar \omega }-1)^{-1}$ is the Bose--Einstein distribution function.

\subsection{Hierarchical equations of motion}
\label{sec:heom}
To construct the HEOM, we need to express $\mathcal{S}(t)$ and $\mathcal{A}(t)$ as linear combinations of the \textit{time basis functions}, $\bm{\phi }(t)=(\dots ,\phi _{k}(t),\dots )^{\mathrm{T}}$, which satisfy the set of time evolution equations
\begin{align}
  \partial _{t}\bm{\phi }(t)=-\bm{\gamma }\bm{\phi }(t).
  \label{eq:basis-evolution}
\end{align}
Here, the superscripts ${}^{\mathrm{T}}$ on the matrices/vectors represent their transposes.
When the column vector $\bm{\phi }(t)$ has $K$ components, $(\bm{\gamma })_{jk}=\gamma _{jk}$ is a $K{\times }K$ matrix.
We assume that $\mathcal{S}(t)$ and $\mathcal{A}(t)$ can be evaluated as
\begin{subequations}
  \begin{align}
    &\begin{aligned}
       \mathcal{S}(t)&=\bm{\sigma }^{\mathrm{T}}\bm{S}\bm{\phi }(t)+s_{\delta }\cdot 2\delta (t)
       \label{eq:S-expansion}
     \end{aligned}
    \intertext{and}
    &\begin{aligned}
       \mathcal{A}(t)&=\bm{\sigma }^{\mathrm{T}}\bm{A}\bm{\phi }(t).
       \label{eq:A-expansion}
     \end{aligned}
  \end{align}
\end{subequations}
Here, $s_{\delta }$ is a real constant, $\bm{\sigma }=(\dots ,\sigma _{k},\dots )^{\mathrm{T}}$ is a constant vector common in $\mathcal{S}(t)$ and $\mathcal{A}(t)$, while $\bm{S}$ and $\bm{A}$ are $K{\times }K$ matrices that commute with $\bm{\gamma }$.

Under this parametrization of the bath correlation functions, the time evolution of the reduced density operator $\hat{\rho }(t)$ is rigorously expressed in terms of a set of time differential equations in the HEOM form as \cite{ikeda2020jcp}
\begin{align}
  \begin{split}
    \partial _{t}\hat{\rho }_{\bm{n}}(t)&=-(\mathcal{L}+\hat{\Xi })\hat{\rho }_{\bm{n}}(t)-\sum _{j,k}n_{j}\gamma _{jk}\hat{\rho }_{\bm{n}-\bm{e}_{j}+\bm{e}_{k}}(t)\\
    &\quad -\sum _{k}\hat{\Phi }_{k}\hat{\rho }_{\bm{n}+\bm{e}_{k}}(t)-\sum _{k}n_{k}\hat{\Theta }_{k}\hat{\rho }_{\bm{n}-\bm{e}_{k}}(t).
    \label{eq:heom}
  \end{split}
\end{align}
Here,
\begin{subequations}
  \begin{align}
    \mathcal{L}&\equiv -\frac{i}{\hbar }\hat{H}^{\times },\quad \hat{\Phi }\equiv \frac{i}{\hbar }\hat{V}^{\times },\quad \hat{\Psi }\equiv \frac{1}{\hbar }\hat{V}^{\circ }\\
    \hat{\bm{\Phi }}&=(\dots ,\hat{\Phi }_{k},\dots )\equiv \bm{\sigma }\hat{\Phi },\\
    \hat{\bm{\Theta }}&=(\dots ,\hat{\Theta }_{k},\dots )\equiv \bm{S}\bm{\phi }(0)\hat{\Phi }-\bm{A}\bm{\phi }(0)\hat{\Psi },\\
    \intertext{and}
    \hat{\Xi }&\equiv -s_{\delta }\hat{\Phi }^{2}.
  \end{align}
\end{subequations}
We have introduced commutation and anti-commutation hyperoperators as $\hat{A}^{\times /\circ }\hat{B}\equiv \hat{A}\hat{B}\mp \hat{B}\hat{A}$.
The multi-index $\bm{n}\equiv (\dots ,n_{k},\dots )$ whose components are non-negative integers represent the index of the ``hierarchy,'' and $\bm{e}_{k}\equiv (0,\dots ,1,0,\dots )$ is the $k$th unit vector.
The first hierarchical element of $\hat{\rho }_{\bm{0}}(t)$ corresponds to the reduced density operator, $\hat{\rho }(t)$, and the remaining hierarchical elements are introduced to facilitate a rigorous treatment of the non-Markovian time dependencies of the bath correlation functions.
Hereafter, we refer to the elements of the hierarchy of $\hat{\rho }_{\bm{n}}(t)$ as the auxiliary density operators (ADOs).
Under the factorized initial condition of Eq.~\eqref{eq:fact-init}, we have $\hat{\rho }_{\bm{0}}(t_{0})=\hat{\rho }(t_{0})$ and $\hat{\rho }_{\bm{n}}(t_{0})=0$ for $\bm{n}\neq \bm{0}$.

\subsection{``Basis'' in hierarchy space}
To simplify the following transformation of HEOM theory, we introduce a series of functions $\psi _{\bm{n}}(\bm{x})$ and their raising and lowering operators $\bm{h}^{+}\equiv (\dots ,h_{k}^{+},\dots )^{\mathrm{T}}$ and $\bm{h}^{-}\equiv (\dots ,h_{k}^{-},\dots )^{\mathrm{T}}$ as
\begin{subequations}
  \begin{align}
    \psi _{\bm{0}}(\bm{x})&\equiv \frac{1}{\sqrt {\mathstrut \pi \left|\bm{D}\right|}}\exp \left(-\bm{x}^{\mathrm{T}}\frac{1}{\bm{D}}\bm{x}\right),\\
    \psi _{\bm{n}}(\bm{x})&\equiv \frac{{\bm{h}^{+}}^{\bm{n}}}{\bm{n}!}\psi _{\bm{0}}(\bm{x}),&&(\bm{n}\neq \bm{0})
  \end{align}
  \label{eq:basis}
\end{subequations}
\begin{subequations}
  \begin{align}
    \bm{h}^{+}&\equiv -\bm{B}^{\mathrm{T}}\sqrt {\mathstrut \frac{\bm{D}}{2}}\bm{\partial }_{x}
    \label{eq:raising},
    \intertext{and}
    \bm{h}^{-}&\equiv \bm{B}^{-1}\left(\sqrt {\frac{2}{\bm{D}}\mathstrut }\bm{x}+\sqrt {\mathstrut \frac{\bm{D}}{2}}\bm{\partial }_{x}\right)
    \label{eq:lowering}.
  \end{align}
\end{subequations}
Here, $\bm{x}=(\dots ,x_{k},\dots )^{\mathrm{T}}$ and $\bm{\partial }_{x}=(\dots ,\partial _{x_{k}},\dots )^{\mathrm{T}}$ are a real vector and its differential operator, respectively, $\bm{D}$ is an arbitrary $K{\times }K$ real positive-definite matrix, and $\bm{B}$ is an arbitrary $K{\times }K$ real regular matrix. The fraction $1/\bm{A}$ denotes the inverse of the matrix $\bm{A}$.
We introduce the following multi-index notation:
\begin{align}
  \begin{aligned}
    \left|\bm{n}\right|&=\sum _{k}n_{k},&&&\bm{n}!&\equiv \prod _{k}n_{k}!,\\
    \bm{\alpha }^{\bm{n}}&\equiv \prod _{k}\alpha _{k}^{n_{k}},&\text{and}&&\delta _{\bm{n},\bm{n}'}&\equiv \prod _{k}\delta _{n_{k},n'_{k}},
  \end{aligned}
\end{align}
where $\delta _{j,k}$ is the Kronecker delta.

Equations~\eqref{eq:raising} and \eqref{eq:lowering} satisfy
\begin{subequations}
  \begin{align}
    \left[h_{j}^{+},h_{k}^{-}\right]&=-\delta _{j,k}
    \intertext{and}
    h_{k}^{-}\psi _{\bm{0}}(\bm{x})&=0.
  \end{align}
\end{subequations}
Therefore, they have the raising and lowering relations of
\begin{subequations}
  \begin{align}
    h_{k}^{+}\psi _{\bm{n}}(\bm{x})&=(n_{k}+1)\psi _{\bm{n}+\bm{e}_{k}}(\bm{x}),\\
    \intertext{and}
    h_{k}^{-}\psi _{\bm{n}}(\bm{x})&=\psi _{\bm{n}-\bm{e}_{k}}(\bm{x}).
  \end{align}
\end{subequations}
These indicate that $\psi _{\bm{n}}(\bm{x})$ pertains to a multi-dimensional Hermite function.
See Appendix \ref{sec:mshf} for more details.

It is noted that $\psi _{\bm{n}}(\bm{x})$ has an orthogonality given by
\begin{align}
  \int \!d\bm{x}\,\psi _{\bm{0}}(\bm{x})^{-1}\cdot \psi _{\bm{n}}(\bm{x})\psi _{\bm{n}'}(\bm{x})&=2^{|\bm{n}|}\prod _{k}(\bm{B}^{\mathrm{T}}\bm{B})_{kk}^{n_{k}}\delta _{\bm{n},\bm{n}'}
  \label{eq:orthogonality}.
\end{align}
This leads to
\begin{align}
  \int \!d\bm{x}\,\psi _{\bm{n}}(\bm{x})&=
  \begin{cases}
    1&\bm{n}=\bm{0}\\
    0&\mathrm{otherwise}
  \end{cases}.
  \label{eq:orthogonality-0}
\end{align}

\begin{subequations}
  Using $\psi _{\bm{n}}(\bm{x})$, we introduce the transformation from the set of ADOs $\hat{\rho }_{\bm{n}}$ to the function
  \begin{align}
    \tilde{f}(\bm{x},t)\equiv \sum _{\bm{n}}\hat{\rho }_{\bm{n}}(t)\psi _{\bm{n}}(\bm{x}),
    \label{eq:expansion}
  \end{align}
  and its inverse transformation
  \begin{align}
    \hat{\rho }_{\bm{n}}(t)=\frac{1}{2^{|\bm{n}|}\prod _{k}(\bm{B}^{\mathrm{T}}\bm{B})_{kk}^{n_{k}}}\int \!d\bm{x}\,\psi _{\bm{0}}(\bm{x})^{-1}\cdot \psi _{\bm{n}}(\bm{x})\tilde{f}(\bm{x},t).
    \label{eq:inverse-expansion}
  \end{align}
\end{subequations}
That is, we regard the ADOs as linear basis coefficients of $\hat{f}(\bm{x},t)$.
The inverse transformation of Eq.~\eqref{eq:inverse-expansion} is a generalization for a construction of the so-called Brinkman hierarchy, which is introduced to express momentum degrees of freedom in the Fokker--Planck--Kramers equation.\cite{risken1989book} 
Similar transformations appear in Refs.~\onlinecite{tanimura1989jpsj, shi2009jcp2, ikeda2015jcp, ikeda2018cp, ikeda2019jctc, yan2020jcp}.

Substituting Eq.~\eqref{eq:expansion} into Eq.~\eqref{eq:heom} provides the time evolution equation of
\begin{align}
  \partial _{t}\tilde{f}(\bm{x},t)&=-\tilde{\mathcal{L}}^{f}(t)\tilde{f}(\bm{x},t),
  \label{eq:fpheom-diagonalized}
\end{align}
where
\begin{align}
  \tilde{\mathcal{L}}^{f}(t)&\equiv \mathcal{L}(t)+\hat{\Xi }+\bm{h}^{+\mathrm{T}}\bm{\gamma }\bm{h}^{-}+\hat{\bm{\Phi }}^{\mathrm{T}}\bm{h}^{-}+\bm{h}^{+\mathrm{T}}\hat{\bm{\Theta }}.
\end{align}
Thus, the discrete degrees of freedom $\bm{n}$ in HEOM theory are mapped into the continuous degrees of freedom $\bm{x}$.
Hereafter, we refer to $\bm{x}$ as the auxiliary modes.

\subsection{Mapping into the collective bath coordinate space}
The right-hand side of Eq.~\eqref{eq:fpheom-diagonalized} contains up to the second-order differential operator of $\bm{x}$, and Eq.~\eqref{eq:orthogonality-0} forces $\tilde{f}(\bm{x},t)$ to satisfy
\begin{align}
  \int \!d\bm{x}\,\tilde{f}(\bm{x},t)&=\hat{\rho }_{\bm{0}}(t)=\hat{\rho }(t).
\end{align}
Thus, $\tilde{f}(\bm{x},t)$ acts as a quasi-distribution and Eq.~\eqref{eq:fpheom-diagonalized} is regarded as a type of generalization for the quantum Fokker--Planck equation.\cite{caldeira1983pa, tanimura1991pra, tanimura2006jpsj, ikeda2017jcp, ikeda2018cp, ikeda2019jctc, iwamoto2021jce}
However, $\bm{\phi }(t)$ usually contains components from the Bose--Einstein distribution $n^{\mathrm{BE}}(\omega )$, which are fundamentally different from the components caused by the spectral density $\mathcal{J}(\omega )$; the latter has mechanical origin while the former does not.
In Eq.~\eqref{eq:fpheom-diagonalized}, both components are coupled to the quantum states of the system through $\hat{\bm{\Phi }}$ and $\hat{\bm{\Theta }}$, which makes it difficult to interpret the physical meaning of the quasi-distribution $\tilde{f}(\bm{x},t)$.
The following introduces the transformation of $\tilde{f}(\bm{x},t)$, which separates these two different types of degrees of freedom.

\subsubsection{Drude spectral density}
Hereafter, we assume that the Drude spectral density is
\begin{align}
  \mathcal{J}^{\mathrm{D}}(\omega )&\equiv 2\eta \frac{\gamma ^{\mathrm{D}}\omega }{\omega ^{2}+{\gamma ^{\mathrm{D}}}^{2}}
  \label{eq:drude},
\end{align}
which is handled in the original HEOM theory.\cite{tanimura1989jpsj}
We employ the decomposition of the Bose--Einstein distribution as
\begin{align}
  n^{\mathrm{BE}}(\omega )+\frac{1}{2}&=\frac{1}{\beta \hbar \omega }+\sum _{\xi }\frac{2\alpha ^{(\xi )}}{\beta \hbar \nu ^{(\xi )}}\frac{\omega {\nu ^{(\xi )}}^{2m^{(\xi )}-1}}{(\omega ^{2}+{\nu ^{(\xi )}}^{2})^{m^{(\xi )}}}\notag\\
  &\quad +\Delta \beta \hbar \omega .
  \label{eq:BE-dist}
\end{align}
This form unifies the Pade spectral decomposition (PSD) and Fano spectral decomposition (FSD) schemes.\cite{hu2010jcp, hu2011jcp, cui2019jcp}
To obtain $\mathcal{S}(t)$ and $\mathcal{A}(t)$, we evaluate the integrals over poles on $\omega =\pm i{\nu ^{(\xi )}}$.
In previous works, the higher-order poles for $m^{(\xi )}>1$ were evaluated using the residue theorem, which creates complicated expressions for the coefficients and increases the difficulty of analytic transformations.\cite{zhang2020jcp, ikeda2020jcp}
Appendix \ref{sec:mr} introduces a new evaluation theorem for higher-order residues using matrix forms. The results in the present case of the Drude spectral density are
\begin{subequations}
  \begin{align}
    \mathcal{S}(t)&=\sigma ^{\mathrm{D}}S^{\mathrm{D}}\phi ^{\mathrm{D}}(t)+\sum _{\xi }{\bm{\sigma }^{\mathrm{BE}(\xi )}}^{\mathrm{T}}\bm{S}^{\mathrm{BE}(\xi )}\bm{\phi }^{\mathrm{BE}(\xi )}(t)\notag\\
    &\quad +s_{\delta }\cdot 2\delta (t)
    \label{eq:S-matrix-rep}\\
    \intertext{and}
    \mathcal{A}(t)&=\sigma ^{\mathrm{D}}A^{\mathrm{D}}\phi ^{\mathrm{D}}(t)
  \end{align}
  for $t\geq 0$, where
\end{subequations}
\begin{subequations}
  \begin{align}
    \phi ^{\mathrm{D}}(t)&\equiv \gamma _{\mathrm{D}}e^{-\gamma _{\mathrm{D}}t},\\
    S^{\mathrm{D}}&\equiv -2\mathrm{Im}\left\{\eta \left(n_{\mathrm{BE}}(i\gamma ^{\mathrm{D}})+\frac{1}{2}\right)\right\},\\
    A^{\mathrm{D}}&\equiv -\eta ,\\
    \sigma ^{\mathrm{D}}&\equiv 1,
  \end{align}
\end{subequations}
\begin{subequations}
  \begin{align}
    \bm{\phi }^{\mathrm{BE}(\xi )}(t)&\equiv \tilde{\bm{\nu }}^{(\xi )}e^{-\tilde{\bm{\nu }}^{(\xi )}}\bm{e}_{1}^{(\xi )},\\
    \bm{S}^{\mathrm{BE}(\xi )}&\equiv -2\mathrm{Im}\left\{\mathcal{J}^{\mathrm{D}}(i\tilde{\bm{\nu }}^{(\xi )})\frac{2\alpha ^{(\xi )}}{\beta \hbar \nu ^{(\xi )}}\frac{{\nu ^{(\xi )}}^{m^{(\xi )}}}{(\tilde{\bm{\nu }}^{(\xi )}+\nu ^{(\xi )})^{m^{(\xi )}}}\right\},\\
    \bm{\sigma }^{\mathrm{BE}(\xi )}&\equiv \bm{e}_{m_{\xi }}^{(\xi )},
  \end{align}
\end{subequations}
and
\begin{align}
  s_{\delta }&\equiv 2\eta \Delta \beta \hbar \gamma _{\mathrm{D}}.
\end{align}
\begin{subequations}
  Here, $\tilde{\bm{\nu }}^{(\xi )}$ is an $m^{(\xi )}{\times }m^{(\xi )}$ matrix defined as
  \begin{align}
    \tilde{\bm{\nu }}^{(\xi )}&\equiv
    \nu ^{(\xi )}
    \begin{pmatrix}
      1 & 0 & \dots & & 0\\
      -1 & 1 & 0 & \ddots & \\
      0 & \ddots & \ddots & 0 & \vdots \\
      \vdots & \ddots & \ddots & 1 & 0\\
      0 & \dots & 0 & -1 & 1\\
    \end{pmatrix},
  \end{align}
  i.e.,
  \begin{align}
    \left(\tilde{\bm{\nu }}^{(\xi )}\right)_{jk}&=\nu ^{(\xi )}\left(\delta _{j,k}-\delta _{j,k+1}\right).
  \end{align}
\end{subequations}

The $m^{(\xi )}$-vector $\bm{e}_{k}^{(\xi )}$ is the $k$th unit vector ($1\leq k\leq m^{(\xi )}$).
The sum between the matrix $\bm{A}$ and the scalar $B$, $\bm{A}+B$, indicates that $\bm{A}+B\bm{I}$, where $\bm{I}$ is the identity matrix with the same size as $\bm{A}$.
In the above expressions, $\phi ^{\mathrm{D}}(t)$ describes the component from the Drude spectral density, and $\bm{\phi }^{\mathrm{BE}(\xi )}(t)$ represents the component from the $\xi $th pole of Eq.~\eqref{eq:BE-dist}.
The $\bm{\phi }^{\mathrm{BE}(\xi )}(t)$ satisfies $\partial _{t}\bm{\phi }^{\mathrm{BE}(\xi )}(t)=-\tilde{\bm{\nu }}^{(\xi )}\bm{\phi }^{\mathrm{BE}(\xi )}(t)$, and $\bm{s}^{\mathrm{BE}(\xi )}$ commutes with $\tilde{\bm{\nu }}^{(\xi )}$. Therefore, the block-diagonalized expressions are
\begin{widetext}
  \begin{subequations}
    \begin{align}
      \bm{\gamma }&=
      \begin{pmatrix}
        \gamma ^{\mathrm{D}} & & & \bm{0}\\
        & \ddots & & \\
        & & \tilde{\bm{\nu }}^{\mathrm{BE}(\xi )} & \\
        \bm{0} &&&\ddots 
      \end{pmatrix},
      &
      \bm{\phi }(t)&=
      \begin{pmatrix}
        \phi ^{\mathrm{D}}(t)\\
        \vdots \\
        \bm{\phi }^{\mathrm{BE}(\xi )}(t)\\
        \vdots \\
      \end{pmatrix},
      &
      \bm{\sigma }&=
      \begin{pmatrix}
        \sigma ^{\mathrm{D}}\\
        \vdots \\
        \bm{\sigma }^{\mathrm{BE}(\xi )}\\
        \vdots \\
      \end{pmatrix},
      \\
      \bm{S}&=
      \begin{pmatrix}
        S^{\mathrm{D}} & & & \bm{0}\\
        & \ddots & & \\
        & & \bm{S}^{\mathrm{BE}(\xi )} & \\
        \bm{0} &&&\ddots 
      \end{pmatrix},
      &
      &
      \text{and}
      &
      \bm{A}&=
      \begin{pmatrix}
        A^{\mathrm{D}} & & & \bm{0}\\
        & \ddots & & \\
        & & \bm{0} & \\
        \bm{0} &&&\ddots 
      \end{pmatrix}.
    \end{align}
  \end{subequations}
  These satisfy the required conditions to construct HEOM theory.
  The corresponding Eqs.\eqref{eq:heom} and \eqref{eq:fpheom-diagonalized} are
  \begin{align}
    \begin{split}
      \partial _{t}\hat{\rho }_{n^{\mathrm{D}},\bm{n}^{\mathrm{BE}}}(t)&=-\left(\mathcal{L}(t)+\hat{\Xi }\right)\hat{\rho}_{n^{\mathrm{D}},\bm{n}^{\mathrm{BE}}}(t)\\
      &\quad -n^{\mathrm{D}}\gamma^{\mathrm{D}}\hat{\rho }_{n^{\mathrm{D}},\bm{n}^{\mathrm{BE}}}(t)\\
      &\quad-\hat{\Phi }\rho_{n^{\mathrm{D}}+1,\bm{n}^{\mathrm{BE}}}(t)-n^{\mathrm{D}}\hat{\Theta}^{\mathrm{D}}\rho _{n^{\mathrm{D}}-1,\bm{n}^{\mathrm{BE}}}(t)\\
      &\quad -\sum _{\xi }\sum_{j,k}n^{\mathrm{BE}(\xi )}_{j}\tilde{\nu }^{(\xi )}_{jk}\hat{\rho}_{n^{\mathrm{D}},\bm{n}^{\mathrm{BE}}-\bm{e}^{(\xi )}_{j}+\bm{e}^{(\xi )}_{k}}(t)\\
      &\quad-\sum _{\xi }\sum _{k}\hat{\Phi }\bm{\sigma }^{\mathrm{BE}(\xi )}_{k}\hat{\rho }_{n^{\mathrm{D}},\bm{n}^{\mathrm{BE}}+\bm{e}^{(\xi )}_{k}}(t)-\sum _{\xi }\sum_{j}n^{\mathrm{BE}(\xi )}_{j}\hat{\Theta }^{\mathrm{BE}(\xi )}_{j}\hat{\rho }_{n^{\mathrm{D}},\bm{n}^{\mathrm{BE}}-\bm{e}^{(\xi )}_{j}}(t)
      \label{eq:heom-drude}
    \end{split}
  \end{align}
  and
  \begin{align}
    \begin{split}
      \partial _{t}\tilde{f}(x^{\mathrm{D}},\bm{x}^{\mathrm{BE}},t)&=-\tilde{\mathcal{L}}^{f}(t)\tilde{f}(x^{\mathrm{D}},\bm{x}^{\mathrm{BE}},t),
    \end{split}
    \label{eq:fpheom-time-evolution-drude}
  \end{align}
  where
  \begin{align}
    \begin{split}
      \tilde{\mathcal{L}}^{f}(t)&\equiv \mathcal{L}(t)+\hat{\Xi }+{h^{\mathrm{D}}}^{+}\gamma^{\mathrm{D}}{h^{\mathrm{D}}}^{-}+\hat{\Phi }{h^{\mathrm{D}}}^{-}+{h^{\mathrm{D}}}^{+}\hat{\Theta}^{\mathrm{D}}\\
      &\quad \quad +\sum _{\xi }\left({\bm{h}^{\mathrm{BE}(\xi )+\mathrm{T}}}\tilde{\bm{\nu }}^{(\xi )}\bm{h}^{\mathrm{BE}(\xi )-}+\hat{\Phi }{\bm{\sigma }^{\mathrm{BE}(\xi )\mathrm{T}}}\bm{h}^{\mathrm{BE}(\xi )-}+{\bm{h}^{\mathrm{BE}(\xi )+\mathrm{T}}}\hat{\bm{\Theta }}^{\mathrm{BE}(\xi )}\right)
    \end{split}
    \label{eq:fpheom-L-drude}
  \end{align}
\end{widetext}
Here, we introduce the hierarchy multi-index as $\bm{n}=(n^{\mathrm{D}},\bm{n}^{\mathrm{BE}})$, the corresponding continuous degrees of freedom as $\bm{x}=(x^{\mathrm{D}},\bm{x}^{\mathrm{BE}})$, and the raising/lowering operators as $\bm{h}^{\pm }=({h^{\mathrm{D}}}^{\pm },{\bm{h}^{\mathrm{BE}}}^{\pm })$.
The multi-index/vector for the components of the Bose--Einstein distribution consists of the set of sub-multi-indices/vectors for the $\xi $th pole (e.g., $\bm{n}^{\mathrm{BE}}=(\dots ,\bm{n}^{\mathrm{BE}(\xi )},\dots )$, and $\bm{n}^{\mathrm{BE}(\xi )}$ is the $m^{(\xi )}$-dimensional multi-index, e.g., $\bm{n}^{\mathrm{BE}(\xi )}=(n_{0}^{\mathrm{BE}(\xi )},\dots ,n_{m^{(\xi )}}^{\mathrm{BE}(\xi )})$). The summations of $j$ and $k$ run over $j, k = 1, \dots , m^{(\xi )}$.
Thus, $\bm{n}^{\mathrm{BE}}$ has $K^{\mathrm{BE}}$ dimensions, where $K^{\mathrm{BE}}\equiv \sum _{\xi }m^{(\xi )}$.
The operators that appear in these expressions are defined as
\begin{subequations}
  \begin{align}
    \hat{\Theta}^{\mathrm{D}}&\equiv S^{\mathrm{D}}\phi ^{\mathrm{D}}(0)\hat{\Phi }-A^{\mathrm{D}}\phi ^{\mathrm{D}}(0)\hat{\Psi },\\
    \hat{\bm{\Theta }}^{\mathrm{BE}(\xi )}&\equiv \bm{S}^{\mathrm{BE}(\xi )}\bm{\phi }^{\mathrm{BE}(\xi )}(0)\hat{\Phi },\\
    \intertext{and}
    \hat{\Xi }&\equiv -s_{\delta }\hat{\Phi}^{2}.
  \end{align}
\end{subequations}

\subsubsection{Mixing and renormalizing the ``hierarchy space''}
As mentioned above, the auxiliary modes $\bm{x}^{\mathrm{BE}(\xi )}$ that come from Eq.~\eqref{eq:BE-dist} are coupled to system states through the operators $\hat{\Phi }\bm{h}^{\mathrm{BE}(\xi )\pm }$, which makes it difficult to interpret the roles of $x^{\mathrm{D}}$ and $\bm{x}^{\mathrm{BE}}$.
Here, we consider the transformation
\begin{align}
  \hat{f}\left(x^{\mathrm{D}},\bm{x}^{\mathrm{BE}},t\right)&\equiv \mathcal{S}^{\mathrm{ren}}\mathcal{S}^{\mathrm{mix}}\tilde{f}\left(x^{\mathrm{D}},\bm{x}^{\mathrm{BE}},t\right)
  \label{eq:hrchy-transform},
\end{align}
\begin{widetext}
  \noindent where
  \begin{align}
    \mathcal{S}^{\mathrm{mix}}&\equiv \exp \left[\frac{a^{\mathrm{D}}}{2}\left(h^{\mathrm{D}+}\right)^{2}+\sum _{\xi }\left({\bm{h}^{\mathrm{BE}(\xi )+\mathrm{T}}}\bm{a}^{\mathrm{BE}(\xi )+}+{\bm{a}^{\mathrm{BE}(\xi )-\mathrm{T}}}\bm{h}^{\mathrm{BE}(\xi )-}\right)h^{\mathrm{D}+}\right]
    \label{eq:hrchy-transform-mix}\\
    \intertext{and}
    \mathcal{S}^{\mathrm{ren}}&\equiv \exp \left[\sum _{\xi }{\bm{h}^{\mathrm{BE}(\xi )+\mathrm{T}}}\ln \left(-\bm{C}^{\mathrm{BE}(\xi )}\right){\bm{h}^{\mathrm{BE}(\xi )-}}\right].
    \label{eq:hrchy-transform-ren}
  \end{align}
  Here, $a^{\mathrm{D}}$ is a constant, $\bm{a}^{\mathrm{BE}(\xi )\pm }$ are $m^{(\xi )}$-dimensional vectors, and $\bm{C}^{\mathrm{BE}(\xi )}$ is an $m^{(\xi )}{\times }m^{(\xi )}$ regular matrix.
  The transformation from Eq.~\eqref{eq:hrchy-transform-mix} causes the modes $x^{\mathrm{D}}$ and $\bm{x}^{\mathrm{BE}}$ to mix in hierarchy space as
  \begin{align}
    \left\{
    \begin{aligned}
      \mathcal{S}^{\mathrm{mix}}h^{\mathrm{D}+}{\mathcal{S}^{\mathrm{mix}}}^{-1}&=h^{\mathrm{D}+},\\
      \mathcal{S}^{\mathrm{mix}}h^{\mathrm{D}-}{\mathcal{S}^{\mathrm{mix}}}^{-1}&=h^{\mathrm{D}-}-\sum _{\xi }\left({\bm{h}^{\mathrm{BE}(\xi )+\mathrm{T}}}\bm{a}^{\mathrm{BE}(\xi )+}+{\bm{a}^{\mathrm{BE}(\xi )-\mathrm{T}}}\bm{h}^{\mathrm{BE}(\xi )-}\right)-a^{\mathrm{D}}h^{\mathrm{D}+},\\
      \mathcal{S}^{\mathrm{mix}}\bm{h}^{\mathrm{BE}(\xi )+\mathrm{T}}{\mathcal{S}^{\mathrm{mix}}}^{-1}&=\bm{h}^{\mathrm{BE}(\xi )+\mathrm{T}}+\bm{a}^{\mathrm{BE}(\xi )-\mathrm{T}}h^{\mathrm{D}+},\\
      \mathcal{S}^{\mathrm{mix}}\bm{h}^{\mathrm{BE}(\xi )-}{\mathcal{S}^{\mathrm{mix}}}^{-1}&=\bm{h}^{\mathrm{BE}(\xi )-}-\bm{a}^{\mathrm{BE}(\xi )+}h^{\mathrm{D}+}.
    \end{aligned}
    \right.
  \end{align}
  On the other hand, the transformation from Eq.~\eqref{eq:hrchy-transform-ren} causes renormalization of the raising/lowering relations in $\bm{x}^{\mathrm{BE}}$ as
  \begin{align}
    \left\{
    \begin{aligned}
      \mathcal{S}^{\mathrm{ren}}\bm{h}^{\mathrm{BE}(\xi )+\mathrm{T}}{\mathcal{S}^{\mathrm{ren}}}^{-1}&=\bm{h}^{\mathrm{BE}(\xi )+\mathrm{T}}\bm{C}^{\mathrm{BE}(\xi )},\\
      \mathcal{S}^{\mathrm{ren}}\bm{h}^{\mathrm{BE}(\xi )-}{\mathcal{S}^{\mathrm{ren}}}^{-1}&=\bm{C}^{\mathrm{BE}(\xi )-1}\bm{h}^{\mathrm{BE}(\xi )-}.
    \end{aligned}
    \right.
  \end{align}
\end{widetext}
Under this transformation, the time evolution of $\hat{f}(x^{\mathrm{D}},\bm{x}^{\mathrm{BE}},t)$ can be evaluated as
\begin{align}
  \partial _{t}\hat{f}(x^{\mathrm{D}},\bm{x}^{\mathrm{BE}},t)&=-\hat{\mathcal{L}}^{f}(t)\hat{f}\left(x^{\mathrm{D}},\bm{x}^{\mathrm{BE}},t\right),
\end{align}
where
\begin{align}
  \hat{\mathcal{L}}^{f}(t)\equiv \mathcal{S}^{\mathrm{ren}}\mathcal{S}^{\mathrm{mix}}\tilde{\mathcal{L}}^{f}(t){\mathcal{S}^{\mathrm{mix}}}^{-1}{\mathcal{S}^{\mathrm{ren}}}^{-1}.
\end{align}
We choose the parameters for these transformations as
\begin{subequations}
  \begin{align}
    \bm{a}^{\mathrm{BE}(\xi )+}&=\bm{s}^{\mathrm{BE}(\xi )}\bm{\phi }^{\mathrm{BE}(\xi )}(0),\\
    \bm{a}^{\mathrm{BE}(\xi )-}&=\bm{\sigma }^{\mathrm{BE}(\xi )},\\
    a^{\mathrm{D}}&=-\frac{1}{\gamma^{\mathrm{D}}}\sum _{\xi }{\bm{\sigma }^{\mathrm{BE}(\xi )}}^{\mathrm{T}}\tilde{\bm{\nu }}^{(\xi )}\bm{s}^{\mathrm{BE}(\xi )}\bm{\phi }^{\mathrm{BE}(\xi )}(0),
  \end{align}
\end{subequations}
and
\begin{align}
  \bm{C}^{\mathrm{BE}(\xi )}&=(\gamma ^{\mathrm{D}}-\tilde{\bm{\nu }}^{(\xi )})/\tilde{g}.
\end{align}
Here, $\tilde{g}$ is a scaling constant that is determined later.
Then All the terms that include $\hat{\Phi }\bm{h}^{\mathrm{BE}(\xi )\pm }$ in Eq.~\eqref{eq:fpheom-L-drude} cancel and we obtain
\begin{widetext}
  \begin{align}
    \begin{split}
      \hat{\mathcal{L}}^{f}(t)&=\mathcal{L}(t)+\hat{\Xi }+h^{\mathrm{D}+}\gamma^{\mathrm{D}}h^{\mathrm{D}-}+\hat{\Phi }h^{\mathrm{D}-}+h^{\mathrm{D}+}\left(S'^{\mathrm{D}}\hat{\Phi }-A'^{\mathrm{D}}\hat{\Psi }\right)\\
      &\quad \quad +\sum _{\xi }\Biggl(\bm{h}^{\mathrm{BE}(\xi )+\mathrm{T}}\tilde{\bm{\nu }}^{(\xi )}\bm{h}^{\mathrm{BE}(\xi )-}\\
      &\quad \quad \quad \quad \quad +h^{\mathrm{D}+}\bm{h}^{\mathrm{BE}(\xi )+\mathrm{T}}\bm{S}'^{\mathrm{BE}(\xi )}\bm{\phi }^{\mathrm{BE}(\xi )}(0)-h^{\mathrm{D}+}\bm{\sigma }'^{\mathrm{BE}(\xi )\mathrm{T}}\bm{h}^{\mathrm{BE}(\xi )-}\Biggr).
    \end{split}
    \label{eq:fpheom-L}
  \end{align}
\end{widetext}
Here,
\begin{subequations}
  \begin{align}
    S'^{\mathrm{D}}&\equiv 2\eta \left(\frac{1}{\beta \hbar }+\sum _{\xi }\frac{2\alpha ^{(\xi )}}{\beta \hbar }\delta _{1,m^{(\xi )}}-\Delta \beta \hbar {\gamma _{\mathrm{D}}}^{2}\right),
  \end{align}
  \begin{align}
    \bm{S}'^{\mathrm{BE}(\xi )}&\equiv ({\tilde{\bm{\nu }}^{(\xi )}}^{2}-{\gamma ^{\mathrm{D}}}^{2})\bm{S}^{\mathrm{BE}(\xi )}/\tilde{g},
  \end{align}
  \begin{align}
    \bm{\sigma }'^{\mathrm{BE}(\xi )}&\equiv \tilde{g}\bm{\sigma }^{\mathrm{BE}(\xi )},\\
    \intertext{and}
    A'^{\mathrm{D}}&\equiv -\eta {\gamma _{\mathrm{D}}}.
  \end{align}
\end{subequations}
In Eq.~\eqref{eq:fpheom-L}, only $x^{\mathrm{D}}$ couples to the quantum states of the system through $\hat{\Phi }$ and $\hat{\Psi }$.
This indicates that $x^{\mathrm{D}}$ is proportional to the collective bath coordinate $X$.
It is noted that the transformations in Eqs.~\eqref{eq:hrchy-transform-mix} and \eqref{eq:hrchy-transform-ren} do not change the value of the integral
\begin{align}
  &\int \!dx^{\mathrm{D}}\,\int \!d\bm{x}^{\mathrm{BE}}\,\hat{f}\left(x^{\mathrm{D}},\bm{x}^{\mathrm{BE}},t\right)\notag\\
  &=\int \!dx^{\mathrm{D}}\,\int \!d\bm{x}^{\mathrm{BE}}\,\tilde{f}\left(x^{\mathrm{D}},\bm{x}^{\mathrm{BE}},t\right)\notag\\
  &=\hat{\rho }_{\bm{0}}(t)=\hat{\rho }(t).
  \label{eq:reduce}
\end{align}

\subsection{Partial and full quantum Fokker--Planck representation}
To see the role of the mode $x^{\mathrm{D}}$, we set the parameters of the basis $\psi _{\bm{n}}(\bm{x})$ as
\begin{subequations}
  \begin{align}
    \bm{D}&=
    \begin{pmatrix}
      D^{\mathrm{D}} & \bm{0}\\
      \bm{0} & \bm{D}^{\mathrm{BE}}\\
    \end{pmatrix}
    \intertext{and}
    \bm{B}&=
    \begin{pmatrix}
      B^{\mathrm{D}} & \bm{0}\\
      \bm{0} & \bm{B}^{\mathrm{BE}}\\
    \end{pmatrix}.
  \end{align}
\end{subequations}
This leads to
\begin{subequations}
  \begin{align}
    h^{\mathrm{D}+}&=-B^{\mathrm{D}}\sqrt {\mathstrut \mathstrut \frac{\mathstrut D^{\mathrm{D}}}{\mathstrut 2}}\bm{\partial }_{x}
    \label{eq:raising-D}
    \intertext{and}
    h^{\mathrm{D}-}&=+\frac{1}{B^{\mathrm{D}}}\left(\sqrt {\mathstrut \frac{\mathstrut 2}{\mathstrut D^{\mathrm{D}}}}\bm{x}+\sqrt {\mathstrut \frac{\mathstrut D^{\mathrm{D}}}{\mathstrut 2}}\bm{\partial }_{x}\right)
    \label{eq:lowering-D}.
  \end{align}
\end{subequations}
By setting
\begin{subequations}
  \begin{align}
    D^{\mathrm{D}}&=\frac{2}{\omega ^{f}}\left(\frac{1}{\beta \hbar }+\sum _{\xi }\frac{2\alpha ^{(\xi )}}{\beta \hbar }\delta _{1,m^{(\xi )}}-\Delta \beta \hbar {\gamma _{\mathrm{D}}}^{2}\right)\\
    \intertext{and}
    B^{\mathrm{D}}&=-\frac{1}{\sqrt {D^{\mathrm{D}}\eta\mathstrut \omega ^{f}}}
  \end{align}
\end{subequations}
with an arbitrary positive scaling constant $\omega ^{f}$, we obtain
\begin{widetext}
  \begin{align}
    \begin{split}
      \hat{\mathcal{L}}^{f}(t)&=\frac{i}{\hbar }\hat{H}^{f}(t)^{\times }+r^{f}\partial _{q}\left(-\frac{1}{2\hbar }\left(\partial _{q}\hat{H}^{f}(t)\right)^{\circ }\right)+\hat{\Xi }-\left(s_{\delta }^{f}+s_{\Delta }^{f\mathrm{BE}}\right)\partial _{q}^{2}\\
      &\quad +\sum _{\xi }\bm{h}^{\mathrm{BE}(\xi )+\mathrm{T}}\tilde{\bm{\nu }}^{(\xi )}\bm{h}^{\mathrm{BE}(\xi )-}+\partial _{q}\sum _{\xi }\bm{h}^{\mathrm{BE}(\xi )+\mathrm{T}}\bm{s}^{f\mathrm{BE}(\xi )}-\partial _{q}\sum _{\xi }\bm{\sigma }^{\mathrm{BE}(\xi )\mathrm{T}}\bm{h}^{\mathrm{BE}(\xi )-},
    \end{split}
    \label{eq:fpheom-L-drude-b}
  \end{align}
\end{widetext}
where
\begin{align}
  \hat{H}^{f}(t)&\equiv \hat{H}(t)-\hbar g^{f}\hat{V}q+\frac{\hbar \omega ^{f}}{2}q^{2}
  \label{eq:H-ext},\\
  g^{f}&\equiv -\frac{1}{B^{\mathrm{D}}}\sqrt {\mathstrut \frac{2}{D^{\mathrm{D}}}}=\sqrt {\mathstrut 2\eta \mathstrut \omega ^{f}},
\end{align}
\begin{subequations}
  \begin{align}
    r^{f}&\equiv \frac{\gamma ^{\mathrm{D}}}{\omega ^{f}},\\
    s^{f}&\equiv \frac{\gamma^{\mathrm{D}}}{\beta \hbar \omega ^{f}}\left(1+\sum _{\xi }2\alpha ^{(\xi )}\delta _{1,m^{(\xi )}}\right),\\
    \bm{s}^{f\mathrm{BE}(\xi )}&\equiv \bm{S}^{f\mathrm{BE}(\xi )}\bm{\phi }^{\mathrm{BE}(\xi )}(0),\\
    \bm{S}^{f\mathrm{BE}(\xi )}&\equiv \frac{\gamma ^{\mathrm{D}}}{\omega ^{f}}\frac{2\alpha ^{(\xi )}}{\beta \hbar \nu ^{(\xi )}}\frac{2{\tilde{\bm{\nu }}^{(\xi )}}{\nu ^{(\kappa )}}^{m^{(\kappa )}}}{(\tilde{\bm{\nu }}^{(\xi )}+\nu ^{(\xi )})^{m^{(\xi )}}},\\
    s_{\Delta }^{f\mathrm{BE}}&\equiv -\frac{{\gamma ^{\mathrm{D}}}^{2}}{\omega ^{f}}\Delta \beta \hbar \gamma^{\mathrm{D}},
  \end{align}
\end{subequations}
Here, we change the symbol $x^{\mathrm{D}}$ to $q$ and set $\tilde{g}=g^{f}$.
It is noted that in Eq.~\eqref{eq:fpheom-L-drude-b}, the coupling between the system and $q$ is described by and only by an extended Hamiltonian Eq.~\eqref{eq:H-ext}.
Equations~\eqref{eq:H-int} and \eqref{eq:H-ext} give $X=g^{f}q$; hence, $\hat{f}(q,\bm{x}^{\mathrm{BE}},t)$ is a joint distribution between the system space, collective bath coordinate space, and auxiliary modes $\bm{x}^{\mathrm{BE}}$.
It is noted that $q$ entirely describes the non-Markovian part of the system-bath interactions, and the Markovian part caused by the last term of Eq.~\eqref{eq:S-matrix-rep} is described by $\hat{\Xi }$.
It is important that the coupling strength constant $\eta $ appears only in Eq.~\eqref{eq:H-ext} through $g^{f}$; hence, it does not affect the connection between $q$ and $\bm{x}^{\mathrm{BE}}$.

The construction of a joint distribution between the system space and collective bath coordinate space from the ADOs was given by Shi et al.\cite{liu2014jcp}
Our transformation of Eq.~\eqref{eq:hrchy-transform} is regarded as an extension of their work including the auxiliary modes with its time evolution as Eq.~\eqref{eq:fpheom-L-drude-b}.
For details, see Appendix \ref{sec:shi}.
Equation~\eqref{eq:hrchy-transform} is regarded as a special case of the inverse for the transformation of the spectral densities given by Legget.\cite{leggett1984prb, garg1985jcp}
Variants of the transformation are used in the so-called reaction coordinate mapping approaches, which is another series of open quantum theory that describes the non-Markovian, non-perturbative, system-environment coupling effects.\cite{chin2010jmp, martinazzo2011jcp, woods2014jmp}
In our work, a similar mapping is performed in hierarchical space instead of with spectral density functions.

By partly applying Eq.~\eqref{eq:inverse-expansion}, we have
\begin{align}
  \begin{split}
    \hat{f}_{\bm{n}^{\mathrm{BE}}}(q,t)&=\frac{1}{{2^{|\bm{n}^{\mathrm{BE}}|}\prod _{k}({\bm{B}^{\mathrm{BE}}}^{\mathrm{T}}{\bm{B}^{\mathrm{BE}}})_{kk}^{n^{\mathrm{BE}}_{k}}}}\\
    &\quad \times \int \!d\bm{x}^{\mathrm{BE}}\,\psi _{\bm{0}}^{\mathrm{BE}}(\bm{x}^{\mathrm{BE}})^{-1}\psi _{\bm{n}^{\mathrm{BE}}}(\bm{x}^{\mathrm{BE}})\hat{f}(q,\bm{x}^{\mathrm{BE}},t).
  \end{split}
\end{align}
We obtain the partial hierarchical representation (or partial quantum Fokker--Planck representation) of Eq.~\eqref{eq:fpheom-L-drude-b} as
\begin{align}
  \begin{split}
    \partial _{t}\hat{f}_{\bm{n}^{\mathrm{BE}}}(q,t)&=-\left[\frac{i}{\hbar }\hat{H}^{\mathrm{ext}}(t)^{\times }+\Xi \right]\hat{f}_{\bm{n}^{\mathrm{BE}}}(q,t)\\
    &\quad -r^{f}\partial _{q}\left(-\frac{1}{2\hbar }\partial _{q}\hat{H}^{\mathrm{ext}}(t)^{\circ }\right)\hat{f}_{\bm{n}^{\mathrm{BE}}}(q,t)\\
    &\quad +\left(s^{f}+s_{\Delta }^{f\mathrm{BE}}\right)\partial _{q}^{2}\hat{f}_{\bm{n}^{\mathrm{BE}}}(q,t)\\
    &\quad -\sum _{\xi }\sum_{j,k}\tilde{\nu }^{(\xi )}_{jk}n^{\mathrm{BE}(\xi )}_{j}\hat{f}_{\bm{n}^{\mathrm{BE}}-\bm{e}^{(\xi )}_{j}+\bm{e}^{(\xi )}_{k}}(q,t)\\
    &\quad -\sum _{\xi }\sum_{j}\partial _{q}s^{f\mathrm{BE}(\xi )}_{j}n^{\mathrm{BE}(\xi )}_{j}\hat{f}_{\bm{n}^{\mathrm{BE}}-\bm{e}^{(\xi )}_{j}}(q,t)\\
    &\quad +\sum _{\xi }\sum_{j}\partial _{q}\sigma ^{\mathrm{BE}(\xi )}_{j}\hat{f}_{\bm{n}^{\mathrm{BE}}+\bm{e}^{(k)}_{j}}(q,t).
  \end{split}
  \label{eq:partial-h-fpheom}
\end{align}
This is an extension of the multi-state low-temperature quantum Smoluchowski equation (MS-LT-QSE) we derived in previous works.\cite{ikeda2019jctc, ikeda2019jcp}
If the Markovian part from Eq.~\eqref{eq:BE-dist} does not exist (i.e.,~ $\Delta =0$) and we use only poles of order $1$ (i.e.,~$m^{(\xi )}=1$), Eq.~\eqref{eq:partial-h-fpheom} reduces to the MS-LT-QSE.

It is also possible to construct a definite full quantum Fokker--Planck representation by setting the matrix values of $\bm{D}^{\mathrm{BE}}$ and $\bm{B}^{\mathrm{BE}}$.
The following choice creates a simplified equation, and we employ the block diagonalized form of
\begin{subequations}
  \begin{align}
    \bm{D}^{\mathrm{BE}}&=
    \begin{pmatrix}
      \ddots & & \bm{0}\\
      & \bm{D}^{\mathrm{BE}(\xi )} & \\
      \bm{0} & &\ddots 
    \end{pmatrix}
    \intertext{and}
    \bm{B}^{\mathrm{BE}}&=
    \begin{pmatrix}
      \ddots & & \bm{0}\\
      & \bm{B}^{\mathrm{BE}(\xi )} & \\
      \bm{0} & &\ddots 
    \end{pmatrix}.
  \end{align}
\end{subequations}
We set $\bm{D}^{\mathrm{BE}(\xi )}$ as the solution of the following continuous Lyapunov equation \cite{horn1991book}
\begin{align}
  \frac{1}{4}\tilde{\bm{\nu }}^{(\xi )}\bm{D}^{\mathrm{BE}(\xi )}+\bm{D}^{\mathrm{BE}(\xi )}\frac{1}{4}{\tilde{\bm{\nu }}^{(\xi )}}^{\mathrm{T}}&=\bm{\Sigma }^{\mathrm{BE}(\xi )}.
\end{align}
We then choose $\bm{B}^{\mathrm{BE}(\xi )}$ as
\begin{align}
  \bm{B}^{\mathrm{BE}(\xi )}&=-\sqrt {\mathstrut \frac{2\omega ^{\mathrm{D}}}{\gamma ^{\mathrm{D}}}}\bm{D}^{\mathrm{BE}(\xi )-1/2}{\tilde{\bm{\nu }}^{(\xi )-1}}.
\end{align}
This gives
\begin{align}
  \begin{split}
    \hat{\mathcal{L}}^{f}(t)&\equiv \frac{i}{\hbar }\hat{H}^{\mathrm{ext}}(t)^{\times }+r^{f}\left(-\partial _{q}\frac{1}{2\hbar }\hat{H}^{\mathrm{ext}}(t)^{\circ }\right)\\
    &\quad +\hat{\Xi }-\partial _{q}\left(s^{f}+s_{\Delta }^{f\mathrm{BE}}\right)\partial _{q}^{2}\\
    &\quad -\sum _{\xi }\bm{\partial }_{x}^{\mathrm{BE}(\xi )\mathrm{T}}\left(\tilde{\bm{\nu }}^{(\xi )}\bm{x}^{\mathrm{BE}(\xi )}+\bm{\Sigma }^{\mathrm{BE}(\xi )}\bm{\partial }_{x}^{\mathrm{BE}(\xi )}\right)\\
    &\quad +g^{\mathrm{BE}}\sum _{\xi }\partial _{q}\bm{\sigma }^{\mathrm{BE}(\xi )\mathrm{T}}\left(\tilde{\bm{\nu }}^{(\xi )}\bm{x}^{\mathrm{BE}(\xi )}+2\bm{\Sigma }^{\mathrm{BE}(\xi )}\bm{\partial }_{x}^{\mathrm{BE}(\xi )}\right)
    \label{eq:full-fpheom}.
  \end{split}
\end{align}
Here,
\begin{align}
  g^{\mathrm{BE}}&\equiv \sqrt {\mathstrut r^{\mathrm{f}}}\\
  \intertext{and}
  \bm{\Sigma }^{\mathrm{BE}(\xi )}&\equiv \frac{2\alpha ^{(\xi )}}{\beta \hbar }\bm{e}_{1}\bm{e}_{1}^{\mathrm{T}}.
\end{align}

\subsection{Factorized initial condition}
As explained in section~\ref{sec:heom}, for the factorized initial conditions of Eq.~\eqref{eq:fact-init}, only the first hierarchy is non-zero, $\hat{\rho }_{\bm{0}}(t_{0})=\hat{\rho }(t_{0})$.
Hence, the corresponding $\hat{f}(q,\bm{x}^{\mathrm{Be}},t_{0})$ is
\begin{align}
  \hat{f}\left(q,\bm{x}^{\mathrm{BE}},t_{0}\right)&=\mathcal{S}^{\mathrm{ren}}\mathcal{S}^{\mathrm{mix}}\hat{\rho }(t_{0})\psi _{\bm{0}}(q,\bm{x}^{\mathrm{BE}}).
\end{align}
In the case of the partial hierarchical representation Eq.~\eqref{eq:partial-h-fpheom}, this expression causes the following recurrence relations at $t=t_{0}$:
\begin{subequations}
  \begin{align}
    \hat{f}_{\bm{0}}(q,t_{0})&=\frac{1}{\sqrt {2\mathstrut \pi D^{f\mathrm{BE}}_{[\gamma ^{\mathrm{D}}]}}}\exp \left(-\frac{q^{2}}{2D^{f\mathrm{BE}}}\right)
    \label{eq:factorized-initial-partial-f-1}\\
    \hat{f}_{\bm{n}+\bm{e}^{(\xi )}_{k}}(q,t_{0})&=-s^{f\mathrm{BE}(\xi )}_{[\gamma ^{\mathrm{D}}]k}\partial _{q}\hat{f}_{\bm{n}}(q,t_{0})
    \label{eq:factorized-initial-partial-f-2}\\
    \intertext{and}
    \left(q+\frac{D^{f\mathrm{BE}}_{[\gamma ^{\mathrm{D}}]}}{4}\partial _{q}\right)\hat{f}_{\bm{n}}(q,t_{0})&=\sum _{\xi }\sum _{k}s^{f\mathrm{BE}(\xi )}_{[\gamma ^{\mathrm{D}}]k}n^{\mathrm{BE}(\xi )}_{k}\hat{f}_{\bm{n}-\bm{e}^{(\xi )}_{k}}(q,t_{0}),
    \label{eq:factorized-initial-partial-f-3}
  \end{align}
\end{subequations}
where
\begin{align}
  \bm{s}^{f\mathrm{BE}(\xi )}_{[\gamma ^{\mathrm{D}}]}&\equiv \frac{1}{{\tilde{\bm{\nu }}^{(\xi )}}+\gamma ^{\mathrm{D}}}\bm{s}^{f\mathrm{BE}(\xi )}\\
  \intertext{and}
  D^{f\mathrm{BE}}_{[\gamma ^{\mathrm{D}}]}&\equiv \frac{1}{\gamma ^{\mathrm{D}}}\left(s^{f}+s_{\Delta }^{f\mathrm{BE}}-\sum _{\xi }\bm{\sigma }^{\mathrm{BE}(\xi )\mathrm{T}}\bm{s}^{f\mathrm{BE}(\xi )}_{[\gamma ^{\mathrm{D}}]}\right)
  \label{eq:D-variance}.
\end{align}
By substituting Eqs.~\eqref{eq:factorized-initial-partial-f-1}--\eqref{eq:factorized-initial-partial-f-3}, the right-hand-side of Eq.~\eqref{eq:partial-h-fpheom} becomes zero when $\eta =0$ and $H^{\times }\hat{\rho }(t_{0})=0$.
Thus, Eqs.~\eqref{eq:factorized-initial-partial-f-1}--\eqref{eq:factorized-initial-partial-f-3} represents the thermal equilibrium state of the bath as uncoupled to the system.

It is noted that while $D^{f\mathrm{BE}}_{[\gamma ^{\mathrm{D}}]}$ corresponds to the variance of the collective coordinate $q$ that is uncoupled to the system, this diverges when increasing the number of poles of order 1 in the Bose--Einstein distribution decomposition of Eq.~\eqref{eq:BE-dist}.
For the exact Matsubara decomposition with an infinite number of poles,
\begin{align}
  n^{\mathrm{BE}}(\omega )+\frac{1}{2}&=\frac{1}{\beta \hbar \omega }+\sum _{\xi =1}^{\infty }\frac{2}{\beta \hbar }\frac{\omega }{\omega ^{2}+{\nu ^{\mathrm{M}(\xi )}}^{2}},
\end{align}
and we obtain
\begin{align}
  D^{f\mathrm{BE}}_{[\gamma ^{\mathrm{D}}]}&=\sum _{\xi =-\infty }^{\infty }\frac{1}{\beta \hbar \omega ^{f}}\frac{\gamma ^{\mathrm{D}}}{\left|\nu ^{\mathrm{M}(\xi )}\right|+\gamma ^{\mathrm{D}}}\rightarrow \infty .
\end{align}
Here, $\nu ^{\mathrm{M}(\xi )}=2\pi \xi /\beta \hbar $ is the $\xi$th Matsubara frequency.
Hence, we do not quantitatively discuss $\hat{f}\left(q,\bm{x}^{\mathrm{BE}},t\right)$.
However, this gives the exact evaluation of the reduced operator through Eq.~\eqref{eq:reduce}.
The high-frequency components $\nu ^{(\xi )}\gg 1$ do not couple to the quantum states of the system because of the different time scales.
This divergence is a result of the Drude spectral density model of Eq.~\eqref{eq:drude}, which causes $\mathcal{S}(0)\rightarrow \infty $. This comes primarily from the ultraviolet divergence of the corresponding Ohmic spectral density.\cite{ikeda2019jctc}
While the values of $\hat{f}\left(q,\bm{x}^{\mathrm{BE}},t\right)$ may not be interpreted quantitatively, qualitatively considering the values could be useful as they are part of the exact description of the effect for the system-bath interactions.

\subsection{Truncation of hierarchy}
The full and partial hierarchical forms, Eqs.~\eqref{eq:heom-drude} and \eqref{eq:partial-h-fpheom}, consist of sets of infinitely many differential equations. Thus, we need to truncate $\bm{n}$ to perform the numerical calculations.
For the ordinary HEOM of Eq.~\eqref{eq:heom-drude}, we usually take a subset that consists of a definite number of hierarchy elements into the numerical calculations, $\bm{n}\in \{\mathcal{N}\}$, and regard the remaining elements as $\hat{\rho }_{\bm{n}}(t)=0$ ($\bm{n}\notin \{\mathcal{N}\} $).
This treatment implies we can assume the values of the deep hierarchy elements are unchanged from those in the factorized initial condition of Eq.~\eqref{eq:fact-init}.
In the case of Eq.~\eqref{eq:partial-h-fpheom}, the hierarchy elements $\hat{f}_{\bm{n}}(q,t)$ are non-zero under the factorized initial condition, which requires a different strategy.
This paper takes a subset of $\hat{f}_{\bm{n}}(q,t)$ ($\bm{n}\in \{\mathcal{N}\}$) into the numerical calculations and assumes that the remaining elements can be evaluated using the recurrence relation under the factorized initial condition as Eq.~\eqref{eq:factorized-initial-partial-f-2}.
It is crucial to validate these truncation schemes by changing the size of the subset to obtain numerically exact results.

The choice of the subset $\{\mathcal{N}\}$ affects the efficiency of the calculations, and many advanced methods have been proposed within the HEOM framework.\cite{shi2009jcp3, hartle2013prb, liu2014jcp, hartle2015prb}
To simplify the theoretical validations, we choose the subset based on the condition that $\bm{n}$ satisfies $\left|\bm{n}\right|\equiv \mathcal{N}\leq \mathcal{N}^{\mathrm{max}}$.\cite{tanimura1994jpsj, ishizaki2005jpsj}
This is one of the most reliable methods to date.
Most existing methods are introduced to decrease the number of elements for numerical efficiency and can sometimes be unstable.
Hereafter, we refer to $\mathcal{N}^{\mathrm{max}}$ as the depth of the hierarchy and the number of elements in the subsets as the size of the hierarchy.

\section{NUMERICAL RESULTS}
\label{sec:results}

As explained in the previous section, performing the calculations based on equations with hierarchical forms requires treating a finite number of elements numerically.
Generally, a stronger system-bath coupling necessitates a larger subset.
This section shows that the size of the subset as required by the ordinary HEOM in Eq.~\eqref{eq:heom-drude} increases rapidly with the coupling strength, which becomes difficult to handle numerically.
In contrast, while Eq.~\eqref{eq:partial-h-fpheom} requires relatively large computational resources to treat continuous degrees of freedom $q$, it is possible to handle strong coupling cases where the HEOM becomes nearly unsolvable.

Hereafter, we employ the dimensionless units $\hbar =1$ and $k_{\mathrm{B}}=1$ for simplicity.
We employ the Dormand–-Prince method to avoid numerical instabilities caused by time integration schemes, which is one of the Runge--Kutta methods where integration errors are adaptively controlled within fourth-order accuracy.\cite{dormand1986jcam}
We show the calculation results for Redfield theory as a comparison.\cite{redfield1965inbook, yang2002cp, ishizaki2009jcp2}
The calculations based on HEOM and Redfield theories were performed using the PyHEOM library.\cite{ikeda2020jcp}
All calculations were performed using double-precision floating-point numbers.

\subsection{Vibronic wavepacket dynamics}
\label{sec:example1}

This section considers a system in which two electronic states are strongly coupled to a single harmonic oscillator.
The bath modes are coupled to the oscillator, which causes dumped vibrations in the time evolution.
The total Hamiltonian is given as
\begin{align}
  \begin{split}
    \hat{H}^{\mathrm{tot}}&=\left(E_{0}+\frac{\hbar \omega ^{\mathrm{vib}}}{2}\left(\hat{z}-d_{0}\right)^{2}\right)|0\rangle \langle 0|\\
    &\quad +\left(E_{1}+\frac{\hbar \omega ^{\mathrm{vib}}}{2}\left(\hat{z}-d_{1}\right)^{2}\right)|1\rangle \langle 1|\\
    &\quad +J\left(|0\rangle \langle 1|+|1\rangle \langle 0|\right)\\
    &\quad +\frac{\hbar \omega ^{\mathrm{vib}}}{2}\hat{p}^{2}+\sum _{\xi }\frac{\hbar \omega _{\xi }}{2}\left[\hat{p}_{\xi }^{2}+\left(\hat{x}_{\xi }-\frac{g_{\xi }}{\omega _{\xi }}\hat{z}\right)^{2}\right].
    \label{eq:hamiltonian-pi}
  \end{split}
\end{align}
Here, $|0\rangle $ and $|1\rangle $ denote the electronic diabatic states, $E_{0}$ and $E_{1}$ are the stable equilibrium energies of each state, and $J$ is the diabatic coupling between the states.
The operators $\hat{z}$ and $\hat{p}$ are the dimensionless coordinate and conjugate momentum of the system oscillator, respectively, and $\omega ^{\mathrm{vib}}$ is its characteristic frequency.
The stable points of $\hat{z}$ in each electronic state are different, $d_{0}$ for $|0\rangle $ and $d_{1}$ for $|1\rangle $. Therefore, the electronic and vibrational states are coupled.
We assume the Drude spectral density is defined as
\begin{align}
  \mathcal{J}(\omega )&=\frac{\zeta }{\omega ^{\mathrm{vib}}}\frac{{\gamma ^{\mathrm{D}}}^{2}}{\omega ^{2}+{\gamma ^{\mathrm{D}}}^{2}}.
\end{align}
Thus, $\eta =\zeta \gamma ^{\mathrm{D}}/2\omega ^{\mathrm{vib}}$, where $\zeta $ is the coupling strength between the system oscillator and the bath, and $1/\gamma ^{\mathrm{D}}$ is the timescale of the bath response function.
In the pure Ohmic limit $\gamma ^{\mathrm{D}}\rightarrow \infty $, the conditions $\zeta <2\omega ^{\mathrm{vib}}$, $\zeta =2\omega ^{\mathrm{vib}}$, and $\zeta >2\omega ^{\mathrm{vib}}$ correspond to the underdamped, critically damped, and overdamped cases of the oscillator, respectively.
This model is typical for many photoisomerization processes, and similar models are used in existing papers.\cite{christensson2012jpcb, scholes2017nature, rafiq2021nc}

Our calculations treat the vibrational states of $\hat{z}$ explicitly using vibrational eigenfunctions on each diabatic state.
The wavefunction of the system is expressed as a linear combination of $|j\rangle {\otimes }|\psi ^{\mathrm{vib}}_{a;j}\rangle $, where $|\psi ^{\mathrm{vib}}_{a;j}\rangle$ is the $a$th vibrational eigenfunction on $|j\rangle $,
\begin{align}
  \langle z|\psi ^{\mathrm{vib}}_{a;j}\rangle&=\frac{1}{\sqrt {\mathstrut 2^{a}a!\sqrt {\mathstrut \pi }}}H_{a}(z-d_{j})\exp \left(-\frac{1}{2}(z-d_{j})^{2}\right),
\end{align}
where $H_{a}(z)=(-1)^{a}e^{z^{2}}\partial _{z}^{a}e^{-z^{2}}$ is the $a$th Hermite polynomial. 
We introduce the ladder operators that act on $|\psi ^{\mathrm{vib}}_{a;j}\rangle$ as
\begin{align}
  \hat{c}_{j}^{+}|\psi ^{\mathrm{vib}}_{a;j}\rangle=\sqrt {\mathstrut a+1}|\psi ^{\mathrm{vib}}_{a;j+1}\rangle\\
  \intertext{and}
  \hat{c}_{j}^{-}|\psi ^{\mathrm{vib}}_{a;j}\rangle=\sqrt {\mathstrut a}|\psi ^{\mathrm{vib}}_{a;j-1}\rangle.
\end{align}
Then, $\hat{z}$ and $\hat{p}$ can be rewritten as
\begin{align}
  \hat{z}&=\frac{1}{\sqrt {\mathstrut 2}}\left(\hat{c}_{j}^{+}+\hat{c}_{j}^{-}\right)+d_{j}\\
  \intertext{and}
  \hat{p}&=\frac{i}{\sqrt {\mathstrut 2}}\left(\hat{c}_{j}^{+}-\hat{c}_{j}^{-}\right).
\end{align}
Thus, the total Hamiltonian is expressed as
\begin{align}
  \hat{H}^{\mathrm{tot}}&=\hat{H}+\hat{H}^{\mathrm{int}}+\hat{H}^{\mathrm{bath}}+\hat{H}^{\mathrm{c}},
  \label{eq:vibronic-hemiltonian}
\end{align}
where
\begin{subequations}
  \begin{align}
    \hat{H}&=\sum _{j=0,1}\left[E_{j}+\hbar \omega ^{\mathrm{vib}}\left(\hat{c}_{j}^{+}\hat{c}_{j}^{-}+\frac{1}{2}\right)\right]|j\rangle \langle j|\notag\\
    &\quad +J\left(|0\rangle \langle 1|+|1\rangle \langle 0|\right),\\
    \hat{H}^{\mathrm{int}}&=-\hat{V}\sum _{\xi }\hbar g_{\xi }\hat{x}_{\xi },\\
    \hat{H}^{\mathrm{bath}}&=\sum _{\xi }\frac{\hbar \omega _{\xi }}{2}\left(\hat{p}_{\xi }^{2}+\hat{x}_{\xi }^{2}\right),\\
    \hat{H}^{\mathrm{c}}&=\sum _{\xi }\frac{\hbar g_{\xi }^{2}}{2\omega _{\xi }}\hat{V}^{2},
  \end{align}
\end{subequations}
and
\begin{align}
  \hat{V}&=\hat{z}=\sum _{j=0,1}\left[\frac{1}{\sqrt {2\mathstrut }}\left(\hat{c}_{j}^{+}+\hat{c}_{j}^{-}\right)+d_{j}\right]|j\rangle \langle j|.
\end{align}
The last term in Eq.~\eqref{eq:vibronic-hemiltonian}, $\hat{H}^{\mathrm{c}}$, is the so-called counter term.\cite{caldeira1983pa}
It is noted that to express the diabatic coupling term $\propto J$ based on the basis set $\{|\psi ^{\mathrm{vib}}_{a;j}\rangle\}$, we need to evaluate the overlap integral $\langle \psi ^{\mathrm{vib}}_{a;0}|\psi ^{\mathrm{vib}}_{b;1}\rangle$, which is expressed as
\begin{align}
  \langle \psi ^{\mathrm{vib}}_{a;j}|\psi ^{\mathrm{vib}}_{b;k}\rangle&=\exp \left(-\frac{d_{jk}^{2}}{4}\right)\notag\\
  &\quad \times \left\{
  \begin{aligned}
    \sqrt {\mathstrut \frac{2^{b}b!}{2^{a}a!}}(-d_{jk})^{a-b}L^{(a-b)}_{b}\left(\frac{d_{jk}^{2}}{2}\right)&&(a\geq b)\\
    \sqrt {\mathstrut \frac{2^{a}a!}{2^{b}b!}}(+d_{jk})^{b-a}L^{(b-a)}_{a}\left(\frac{d_{jk}^{2}}{2}\right)&&(a\leq b)\\
  \end{aligned}
  \right.,
\end{align}
where $d_{jk}\equiv d_{j}-d_{k}$, and we introduce the associated Laguerre polynomials defined as $L_{n}^{(\alpha )}(x)\equiv x^{-\alpha }e^{x}\partial _{x}^{n}(e^{-x}x^{x+\alpha })/n!$.

\begin{figure}
  \centering
  \includegraphics[scale=\SingleColFigScale]{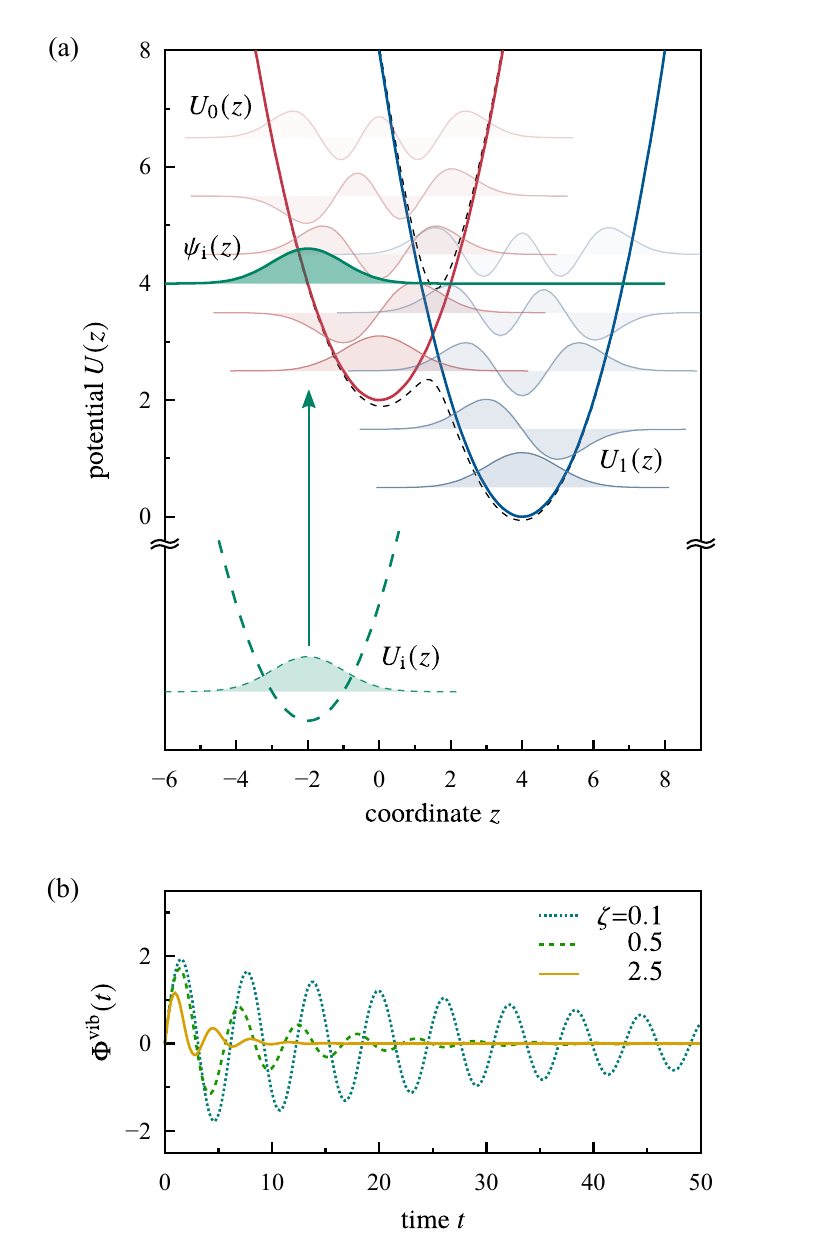}
  \caption{ (a)
    The PESs of our vibronic model system, where the solid red and blue curves represent the diabatic PESs for $|0\rangle $ and $|1\rangle $, respectively.
    Here, $U_{j}(z)=\hbar \omega ^{\mathrm{vib}}(z-d_{j})^{2}/2$ $(j=0,1)$.
    The corresponding adiabatic PESs are depicted as black dashed curves.
    The first five vibrational eigenfunctions on each diabatic surface are also depicted as thin curves.
    The solid green curve represents the initial wavepacket $\psi _{\mathrm{i}}(z)$.
    (b)
    The vibrational response function of the system oscillator $z$ without electronic coupling.
  }
  \label{fig:pot}
\end{figure}
Our calculations utilized the parameters $E_{0}=2$, $E_{1}=0$, $d_{0}=0$, $d_{1}=4$, and $J=0.8$.
We assume the initial wavepacket $\psi _{\mathrm{i}}(z)$ at $t=0$ is the vibrational ground state centered at $d_{\mathrm{i}}=-2$ on $|0\rangle $.
The corresponding diabatic and adiabatic potential energy surfaces (PESs) are illustrated in Fig.~\ref{fig:pot}(a).
The bath parameters are $T=1$ and $\gamma ^{\mathrm{D}}=1$, and the coupling strength $\zeta $ is chosen from $0.1,~0.5,~\text{and}~2.5$.
The time dependencies of the vibrational response function $\Phi ^{\mathrm{vib}}(t)$ under the above conditions are depicted in Fig.~\ref{fig:pot}(b).
It is noted that even in the strong coupling case ($\zeta =2.5$), the oscillator shows underdamped dynamics.

It is important that $\hat{V}=\hat{z}$ is expressed in the basis set $|\psi ^{\mathrm{vib}}_{a;j}\rangle $ as
\begin{align}
  \hat{V}&=\sum _{j=0,1}|\psi ^{\mathrm{vib}}_{a;j}\rangle (\bm{V}_{j})_{ab}\langle \psi ^{\mathrm{vib}}_{b;j}|,
\end{align}
where
\begin{align}
  \bm{V}_{j}&=
  \frac{1}{\sqrt {\mathstrut 2}}
  \begin{pmatrix}
    d_{j} & \sqrt {\mathstrut 1} & 0 & \dots & 0 \\
    \sqrt {\mathstrut 1} & d_{j} & \sqrt {\mathstrut 2} & \ddots & \vdots \\
    0 & \sqrt {\mathstrut 2} & \ddots & \ddots & 0\\
    \vdots & \ddots & \ddots & d_{j} & \sqrt {N^{\mathrm{vib}}{-}1\mathstrut }\\
    0 & \dots & 0 & \sqrt {N^{\mathrm{vib}}{-}1\mathstrut } & d_{j}\\
  \end{pmatrix}.
\end{align}
Here, $N^{\mathrm{vib}}$ is the number of vibrational states on each employed electronic state.
This expression indicates that the effective coupling between the system and the bath is strengthened when employing many vibrational basis functions.
For large $d_{jk}$, the electronic transition between $|j\rangle $ and $|k\rangle $ causes significant vibrational excitations, which requires a large $N^{\mathrm{vib}}$ for an accurate description.
Even for a relatively small system-bath coupling strength ($\zeta$), this $\hat{V}$ can cause a strong non-perturbative regime.

\begin{figure}
  \centering
  \includegraphics[scale=\SingleColFigScale]{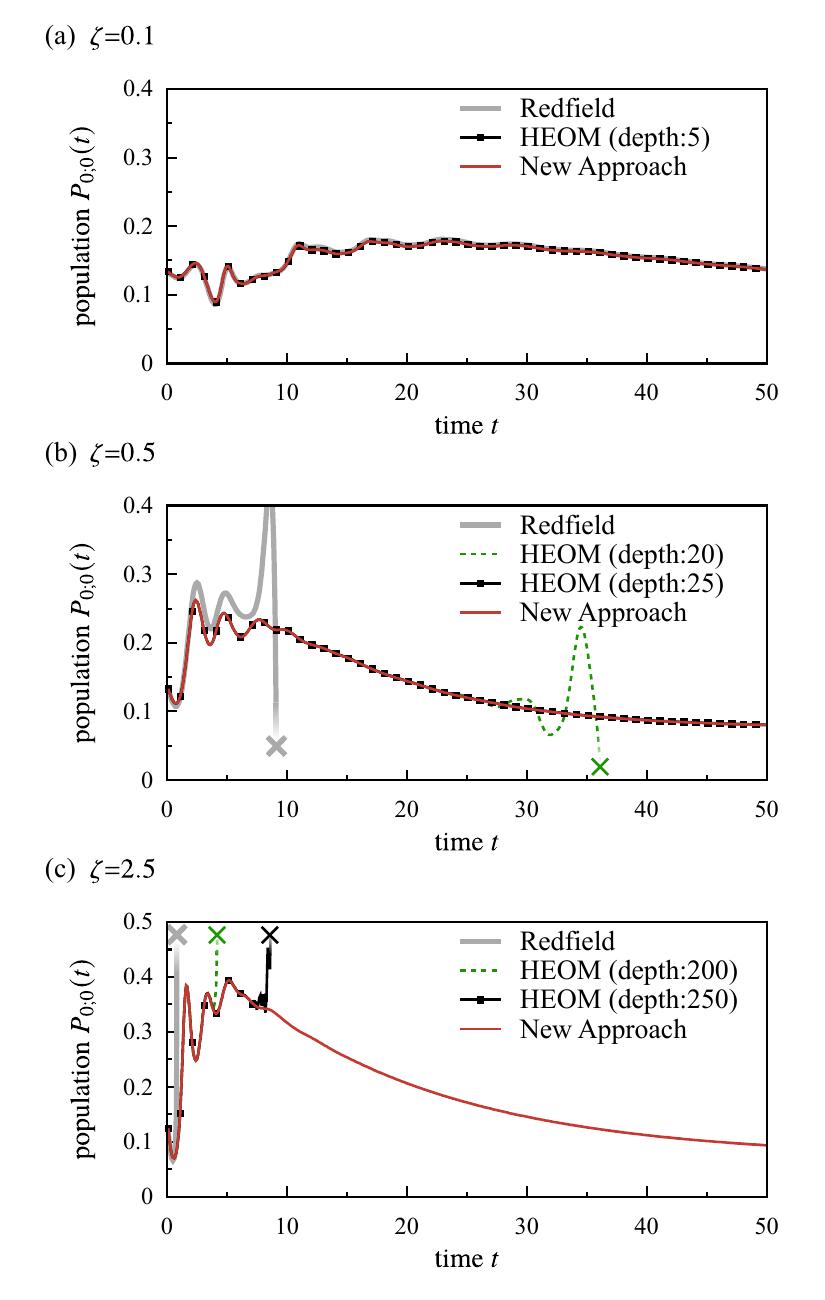}
  \caption{
    Populations of the vibrational ground state on $|0\rangle $ as a function of time $t$ [$P_{0;0}(t)$] for (a) weak, (b) moderate, and (c) strong coupling cases ($\zeta =0.1$, $0.5$, and $2.5$, respectively).
    The red curves represent the results of Eq.~\eqref{eq:partial-h-fpheom}, the green dashed curves and solid black curves with symbols $\blacksquare $ represent those of the HEOM for different $\mathcal{N}^{\mathrm{max}}$, and the gray thick curves indicate those of Redfield theory.
    The $\times$ symbol on each line indicates that the calculation diverges to infinity after the point.
  }
  \label{fig:population}
\end{figure}
In Fig.~\ref{fig:population}, the populations of the vibrational ground state on $|0\rangle $ ($|\psi ^{\mathrm{vib}}_{0;0}\rangle {\otimes }|0\rangle $) as a function of time $t$ are depicted for (a) weak, (b) moderate, and (c) strong coupling cases ($\zeta =0.1$, $0.5$, and $2.5$, respectively).
The simulations were performed using our proposed treatment [Eq.~\eqref{eq:partial-h-fpheom}], the HEOM [Eq.~\eqref{eq:heom-drude}], and the Redfield equations.
We truncate the number of vibrational states on each electronic state by $N^{\mathrm{vib}}=20$ and employ the PSD$[1/1]$ scheme to obtain the parameters for Eq.~\eqref{eq:BE-dist}, which gave $K^{\mathrm{BE}}=1$, $\nu ^{(0)}=6.481$, $\alpha ^{(0)}=1.225$, $m^{(0)}=1$, and $\Delta =0.025$.
In the calculations for Eq.~\eqref{eq:partial-h-fpheom}, we set $\omega ^{f}=1$ and employ a uniform mesh to evaluate the differential operation in the $q$ direction.
The mesh range is $-24\leq q<+24$ with mesh size of $N_{q}=64$.
The finite-difference calculations for the $q$ derivative in Eq.~\eqref{eq:partial-h-fpheom} are performed using the central difference method with sixth-order accuracy.
The values of $\mathcal{N}^{\mathrm{max}}$ for Eqs.~\eqref{eq:partial-h-fpheom} and \eqref{eq:heom-drude} and the required computational memories to treat hierarchical elements are given in Table~\ref{table:cost-1}.
For the Redfield equation, keeping the equation at the second-order in the system-bath interactions requires evaluating the Redfield tensor in the eigenspace of $\hat{H}$ instead of $\hat{H}+\hat{H}^{\mathrm{c}}$.
\begin{table}
  \caption{
    Values of the employed $\mathcal{N}^{\mathrm{max}}$ and the corresponding size of the calculation elements as ratios to the size of the reduced density matrix as $\hat{\rho }(t)$.
  }
  \label{table:cost-1}
  \centering
  \begin{tabular}{crr}
    \hline
    Method & $\mathcal{N}^{\mathrm{max}}$ & Size\\
    \hline \hline
    Proposed Approach & 3~~ & 256 \\
    \hline
    \multirow{5}{*}{HEOM} & 5~~ & 21 \\
    & 20~~ & 231 \\
    & 25~~ & 351 \\
    & 200~~ & 20,301 \\
    & 250~~ & 31,626 \\
    \hline
    Redfield & -- & 1 \\
    \hline
  \end{tabular}
\end{table}

In Fig.~\ref{fig:population}(a), the results for all theories are nearly coincident because the Born and Markov approximations in Redfield theory are valid for sufficiently small coupling.
For moderate coupling, as shown in Fig.~\ref{fig:population}(b), the results of Redfield theory diverge to infinity because the theory is no longer valid.
The results for the HEOM calculations are still valid, but the required size of the hierarchy increases moderately.
For the strong coupling in Fig.~\ref{fig:population}(c), it is difficult to obtain converged results for HEOM theory.
The strong coupling between the system and the bath as caused by $\zeta $ and $\hat{V}$ requires a vast number of hierarchy elements, and deep hierarchy elements require high precision in the numerical calculations. Thus, the coefficients of the second and fifth terms in Eq.~\eqref{eq:heom-drude} become too large.
In such cases, the HEOM is no longer solvable within a fixed precision.
It is noted for HEOM calculations with $\mathcal{N}^{\mathrm{max}}=250$, we employed renormalized ADOs defined as $\hat{\rho }_{\bm{n}}'(t)\equiv \hat{\rho }_{\bm{n}}(t)/\sqrt {\mathstrut \bm{n}!}$ to suppress a floating point overflow error for the deep hierarchy elements.\cite{shi2009jcp3, ikeda2020jcp}
However, this renormalization did not decrease the size of the needed hierarchy.

The proposed treatment in Eq.~\eqref{eq:partial-h-fpheom} gives valid results for all conditions.
This can be interpreted as follows.
In principle, both Eqs.~\eqref{eq:heom-drude} and \eqref{eq:partial-h-fpheom} are equivalent and give the same rigorous results as long as all calculation elements are properly evaluated.
However, when there is a strong coupling between the system and bath, a joint distribution of the system space and collective bath coordinate space becomes significantly distorted.
While Eq.~\eqref{eq:partial-h-fpheom} can directly describe such distortions from the distribution of $\hat{f}(q,\bm{x}^{\mathrm{BE}},t)$, the HEOM in Eq.~\eqref{eq:heom-drude} treats the distortion within the linear basis expansion of Eq.~\eqref{eq:expansion}.
The basis set $\{\psi _{\bm{n}}(\bm{x})\}$ is optimized to express the uncoupled bath equilibrium state; therefore, it can become inefficient when $\hat{f}(q,\bm{x}^{\mathrm{BE}},t)$ is far from the original equilibrium distribution.
In this case, it is more proper to employ the continuous space representation of Eq.~\eqref{eq:partial-h-fpheom} instead of Eq.~\eqref{eq:heom-drude}.
This interpretation also explains why HEOM theory is more efficient than our proposed treatment when the coupling constant is small.
In this case, $\hat{f}(q,\bm{x}^{\mathrm{BE}},t)$ is barely distorted; hence, it can be efficiently expressed using the small basis set in Eq.~\eqref{eq:basis}.

It is also important that our proposed treatment requires only a small hierarchy size with $\{\bm{n}^{\mathrm{BE}}\}$, even in the strong coupling case.
In Eq.~\eqref{eq:partial-h-fpheom}, the coupling strengths of $\eta $ and $\hat{V}$ do not appear in the terms that connect different hierarchy elements.
Hence, the increased coupling strength barely increases the value of $\mathcal{N}^{\mathrm{max}}$ which is required to obtain converged results.

\begin{figure*}
  \centering
  \includegraphics[scale=\DoubleColFigScale]{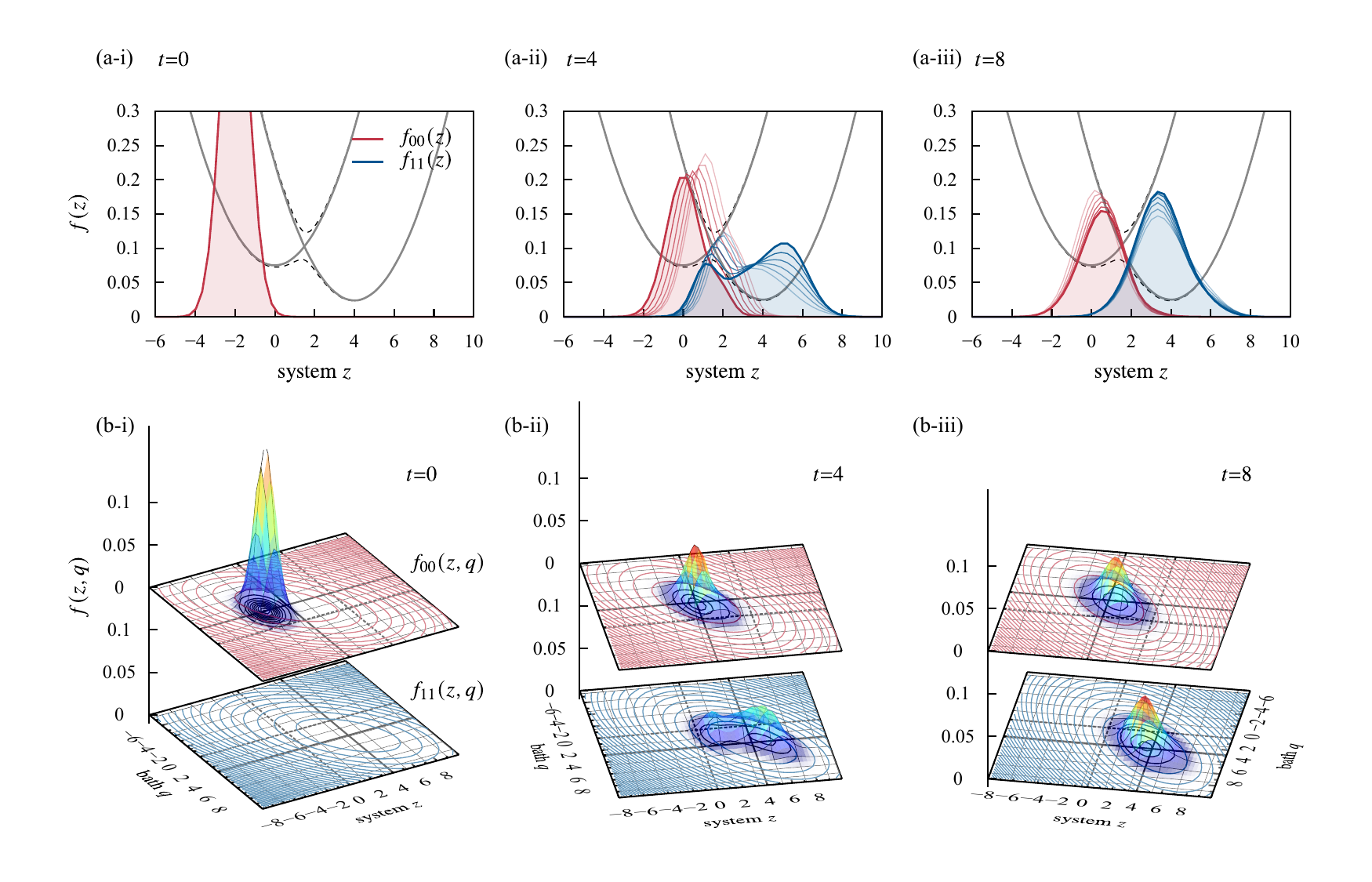}
  \caption{
    (a) Snapshots of wavepackets in system space, $f_{jj}(z,t)$, with waiting times of (i) $t=0$, (ii) $4$, and (iii) $8$.
    The thick red curves and blue curves represent the wavepackets on $|0\rangle $ and $|1\rangle $ at each time, respectively, and the thin curves represent snapshots for a few time steps before, which are drawn to represent the wavepacket motion.
    (b) Snapshots of the wavepackets in the two-dimensional system-bath space, $f_{jj}(z,q,t)$, with waiting times of (i) $t=0$, (ii) $4$, and (iii) $8$.
    The contours of the diabatic PESs are represented by the red and blue contours in the two-dimensional plates for $|0\rangle $ and $|1\rangle $, respectively.
  }
  \label{fig:wavepacket}
\end{figure*}
To examine the physics behind the results in Fig.~\ref{fig:population}, we consider a one-dimensional wavepacket in $z$ space and a two-dimensional wavepacket in $(z,q)$ space defined respectively as
\begin{subequations}
  \begin{align}
    f_{jj}(z,t)&\equiv \int \!dq\,f_{jj}(z,q,t)\\
    \intertext{and}
    f_{jj}(z,q,t)&\equiv \langle z|{\otimes }\langle j|\hat{f}_{\bm{0}}(q,t)|j\rangle {\otimes }|z\rangle.
  \end{align}
\end{subequations}
Figure~\ref{fig:wavepacket} illustrates snapshots of the wavepackets $f_{jj}(z,t)$ and $f_{jj}(z,q,t)$ for the case $\zeta =0.5$.
At $t=0$, the wavepacket is located at $z=d_{\mathrm{i}}=-2$. As we assumed the factorized initial condition, the joint distribution between the system and collective bath coordinate space, $f_{jj}(z,q,0)$, are uncorrelated.
The wavepacket oscillates according to the PES $U_{0}(z)$ and passes the crossing region around $z=z^{\dagger }=1.5$ several times.
This motion causes non-adiabatic (vibronic) transitions; hence, populations on $|0\rangle {\otimes }|\psi ^{\mathrm{vib}}_{0;0}\rangle$ oscillate around $t=4$ in Fig.~\ref{fig:population}(b) and several distinct peaks appear on the state $|1\rangle $ in Figs.~\ref{fig:wavepacket}(a-ii) and \ref{fig:wavepacket}(b-ii).
Here, 
\begin{align}
  z^{\dagger }\equiv \frac{(E_{1}-E_{0})}{\hbar \omega ^{\mathrm{vib}}(d_{1}-d_{0})}+\frac{d_{0}+d_{1}}{2}.
\end{align}
The distribution on the bath coordinate $q$ reorganizes after the non-adiabatic transitions.
For clarity, we plot the diabatic PESs corresponding to the extended Hamiltonian in Eq.~\eqref{eq:H-ext} as contours in Fig.~\ref{fig:wavepacket}(b).
These are defined as
\begin{align}
  U_{j}(z,q)&\equiv E_{j}+\frac{\hbar \omega ^{\mathrm{vib}}}{2}\left(z-d_{j}\right)^{2}-\hbar g^{f}zq+\frac{\hbar \omega ^{f}}{2}q^{2}+H^{\mathrm{c}}\notag\\
  &=E_{j}+\frac{\hbar \omega ^{\mathrm{vib}}}{2}\left(z-d_{j}\right)^{2}+\frac{\hbar \omega ^{f}}{2}\left(q-\frac{g^{f}}{\omega ^{f}}z\right)^{2}.
\end{align}
This extended PES indicates that the $j$th diabatic PESs reorganize the bath coordinate $q$ to $q=d^{q}_{j}$, where
\begin{align}
  d^{q}_{j}&=\frac{g^{f}d_{j}}{\omega ^{f}}=\frac{\sqrt {\mathstrut \gamma ^{\mathrm{D}}\zeta }d_{j}}{\mathstrut \omega ^{f}},
\end{align}
whose value is $d^{q}_{0}=0$ for $|0\rangle $ and $d^{q}_{1}=2\sqrt {\mathstrut 2}\simeq 2.828$ for $|1\rangle $.
After time evolves for a certain period, the oscillation on the state $|0\rangle $ is damped, and the wavepacket on $|0\rangle $ becomes trapped at the minimum on the corresponding adiabatic energy surface [Figs.~\ref{fig:wavepacket}(a-iii) and \ref{fig:wavepacket}(b-iii)].
This trapping causes a gradual time evolution for the populations in Fig.~\ref{fig:population}(b).
Thus, the proposed approach in Eq.~\eqref{eq:partial-h-fpheom} gives robust calculations and bath space information for the system state transitions.

\subsection{Spontaneous de-excitation under near zero-temperature environment}
\label{sec:example3}

As another example, we discuss the spontaneous de-excitation process of a two-level system $\{|g\rangle , |e\rangle \}$ caused by a near zero-temperature bosonic environment.
We assume that the Hamiltonian of the system and the system-bath interaction operator are expressed as $\hat{H}=\hbar \Omega _{e}\hat{c}^{+}\hat{c}^{-}$ and $\hat{V}=\hat{c}^{+}+\hat{c}^{-}$, respectively, where $\hat{c}^{+}\equiv |e\rangle \langle g|$ and $\hat{c}^{-}\equiv |g\rangle \langle e|$ are creation/annihilation operators for the system.
To show that our approach is valid for the FSD scheme, which has several higher-order poles $m^{(\xi )}\geq 2$, we set $T=10^{-2}$ and employ the PSD$[1/1]$+FSD$[6]$ scheme with a reference temperature of $T_{0}=1$, which means that the Bose--Einstein distribution function at $T_{0}$ is evaluated by the PSD$[1/1]$ scheme and the remaining, which is caused by the difference $T-T_{0}$, is evaluated by the FSD scheme, which causes $K^{\mathrm{BE}}=7$.
The result of the parametrization scheme is shown in Table~\ref{table:poles}.

\begin{table}
  \caption{
    Parameters in Eq.~\eqref{eq:BE-dist} employed to evaluate the Bose--Einstein distribution for $T=10^{-2}$.
  }
  \label{table:poles}
  \centering
  \begin{tabular}{c|rrrr}
    \hline & $\xi =0~$ & $1~$ & $2~$ & $3~$ \\
    \hline
    $\nu ^{(\xi )}$ & 6.481~ & 0.7381~ & 0.1815~ & 1.806~ \\
    $\alpha ^{(\xi )}$ & 122.5~ & 49.55~ & 7.990~ & -87.20~ \\
    $m^{(\xi )}$ & 1~ & 1~ & 2~ & 3~\\
    \hline
    $\Delta $ & ~0.00025~ & & &\\
    \hline
  \end{tabular}
\end{table}
\begin{figure}
  \centering
  \includegraphics[scale=\SingleColFigScale]{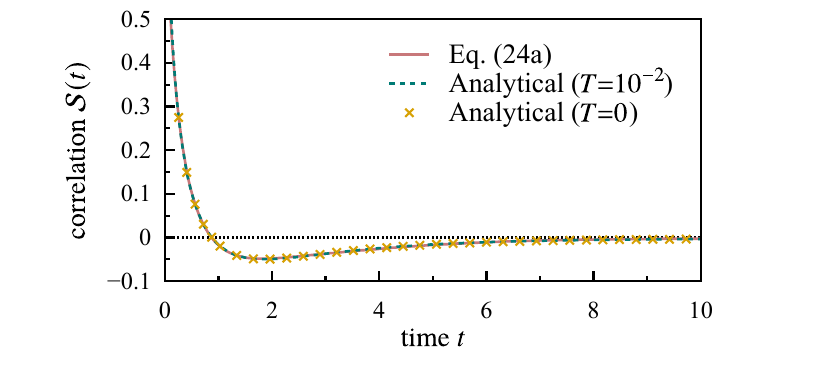}
  \caption{
    Symmetrized correlation function $\mathcal{S}(t)$ for $\zeta =0.5$ and $T=10^{-2}$ with the Drude spectral density from Eq.~\eqref{eq:drude}.
    The solid red and dashed green curves represent Eq.~\eqref{eq:S-matrix-rep} and the analytic evaluation, respectively.
    The $\times $ symbols depicts $\mathcal{S}(t)$ for the zero-temperature limit $T\rightarrow 0$.
  }
  \label{fig:correlation}
\end{figure}
Figure~\eqref{fig:correlation} plots $\mathcal{S}(t)$ to check the accuracy of the expansion and validity of the matrix evaluation for the poles given by Eq.~\eqref{eq:S-matrix-rep}.
We set the initial state of the system as the pure excited state $\hat{\rho }(t_{0})=|e\rangle \langle e|$, and calculate the population dynamics connected to the near zero-temperature bath $T=10^{-2}$.
In this model, transitions between the two states are caused solely by fluctuations in the bath coordinates.
As there are no thermal fluctuations in the zero-temperature limit, the spontaneous transition is dominated by quantum fluctuations from the environment.

In all calculations of Eq.~\eqref{eq:partial-h-fpheom} for this section, we set $\omega ^{f}=1$ and employ a uniform mesh to evaluate differential operation in the $q$ direction. The mesh range is $-12\leq q<+12$ with a mesh size of $N_{q}=128$.
The finite-difference calculations for $q$ derivative in Eq.~\eqref{eq:partial-h-fpheom} are performed using the central difference method with sixth-order accuracy.
As the large constants $\alpha ^{(\xi )}$ in Table~\ref{table:poles} cause strong connections between the hierarchy elements in Eq.~\eqref{eq:partial-h-fpheom}, we need a deeper hierarchy subspace.
The values of $\mathcal{N}^{\mathrm{max}}$ for Eqs.~\eqref{eq:partial-h-fpheom} and \eqref{eq:heom-drude} and the required computational memories to treat hierarchical elements are given in Table~\ref{table:cost-2}.
As the dimension of $\bm{n}^{\mathrm{BE}}$ is seven, increases in $\mathcal{N}^{\mathrm{max}}$ cause a significant increase of the hierarchy size.
\begin{table}
  \caption{
    Values of the employed $\mathcal{N}^{\mathrm{max}}$ and the corresponding size of the calculation elements as ratios to the size of the reduced density matrix as $\hat{\rho }(t)$.
  }
  \label{table:cost-2}
  \centering
  \begin{tabular}{crr}
    \hline
    Method & $\mathcal{N}^{\mathrm{max}}$ & Size\\
    \hline \hline
    Proposed Approach & 9~~ & 1,464,320 
    \\
    \hline
    \multirow{4}{*}{HEOM} & 2~~ & 45 \\
    & 7~~ & 6,435 \\
    & 20~~ & 3,108,105 \\
    & 25~~ & 13,884,156 \\
    \hline
    Redfield & -- & 1 \\
    \hline
  \end{tabular}
\end{table}

\begin{figure}
  \centering
  \includegraphics[scale=\SingleColFigScale]{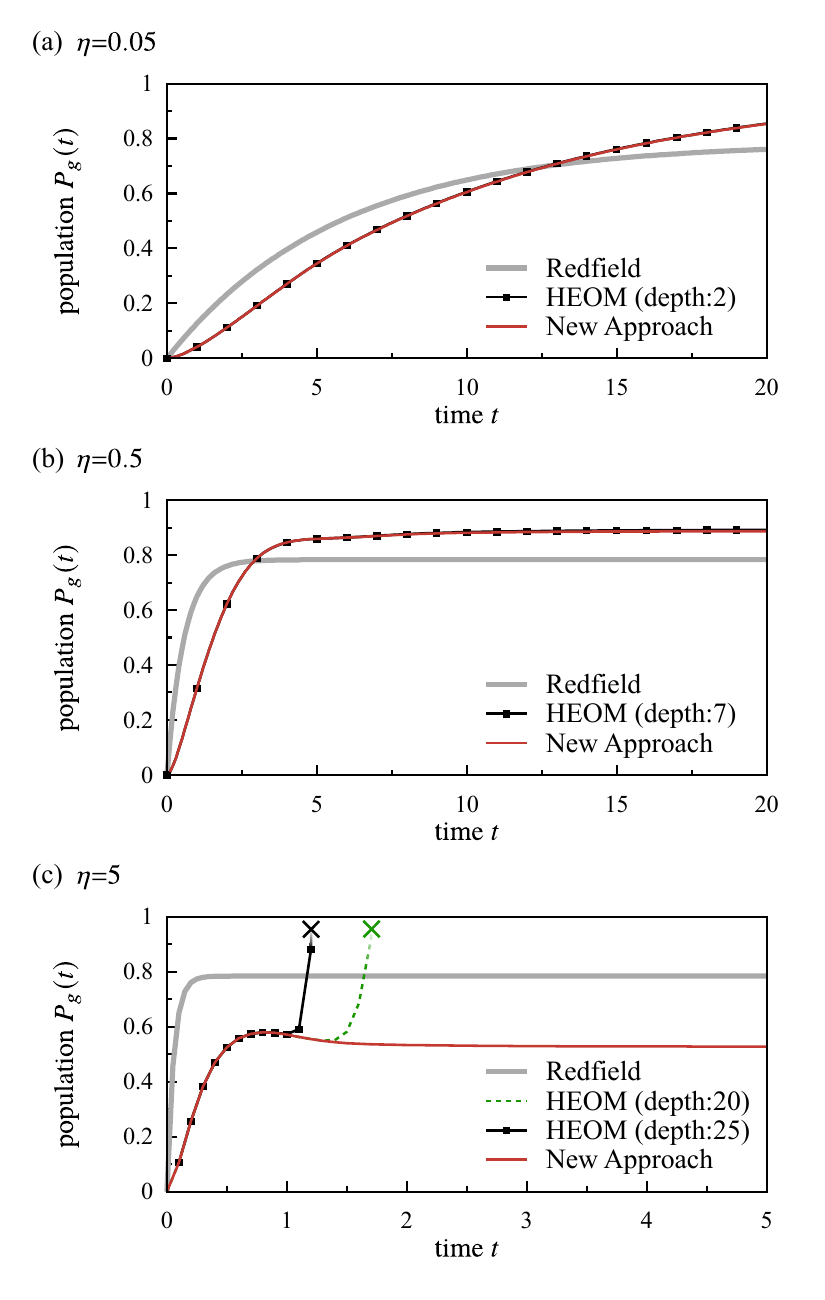}
  \caption{
    Populations of the ground state on $|g\rangle $ as a function of time $t$ as given by $P_{g}(t)$ are depicted for (a) weak, (b) moderate, and (c) strong coupling cases ($\eta =0.05$, $0.5$, and $\eta =5$, respectively).
    The red curves represent the results of Eq.~\eqref{eq:partial-h-fpheom}, the green dashed curves and solid black curves with symbols $\blacksquare $ represent those of HEOM theory with different $\mathcal{N}^{\mathrm{max}}$, and the gray thick curves are those of Redfield theory.
    The symbol $\times $ on each line indicates that the calculation diverged to infinity after the indicated point.
  }
  \label{fig:zerotemp_population}
\end{figure}
Figure~\eqref{fig:zerotemp_population} depicts the numerical results of the spontaneous de-excitation of the excited state population in the case of weak, moderate, and strong coupling cases ($\eta =0.05$, $0.5$, and $5$).
The parameters are $\Omega _{0}=1$ and $\gamma ^{\mathrm{D}}=1$, and the scaling constant $\omega ^{\mathrm{f}}$ in Eq.~\eqref{eq:partial-h-fpheom} is $\omega ^{\mathrm{f}}=1$.
In Fig.~\ref{fig:zerotemp_population}(a), the results of Redfield theory and the other calculations are already different even though it is in a weak coupling regime.
As only $\hat{V}$ causes the spontaneous de-excitation in this model, the non-Markovian nature of the bath correlation functions is more important than in the previous section. Thus, we cannot employ the Markovian approximation, which is included in Redfield theory.\cite{ikeda2020jcp}

In Figs.~\ref{fig:zerotemp_population}(a) and \ref{fig:zerotemp_population}(b), the results of Eq.~\eqref{eq:partial-h-fpheom} agree well with the HEOM results.
However, we could not perform convergent HEOM calculations for the strong coupling case, as seen in Fig.~\ref{fig:zerotemp_population}(c).
While this is similar to the case in Fig.~\ref{fig:population} in the previous section, the required hierarchy of HEOM theory is further numerically demanding because $\bm{n}^{\mathrm{BE}}$ here has a higher dimensionality.
It is noted that the HEOM calculations for a deeper hierarchy ($\mathcal{N}^{\mathrm{max}}=25$) show severe divergence than for a shallower hierarchy ($\mathcal{N}^{\mathrm{max}}=20$) in the strong coupling case.
The deep hierarchy requires high-precision calculations, which has an increased $\mathcal{N}^{\mathrm{max}}$ that could cause unsolvable equations before the results converge.
As discussed in the previous section, our proposed approach in Eq.~\eqref{eq:partial-h-fpheom} gives the exact solution, even if the system-bath coupling strength moves beyond the regime where HEOM theory can practically handle.
The values of the equilibrium population $P_{g}(\infty )$ approaches $0.5$ as the coupling strength $\eta $ increases.
As the system Hamiltonian $\hat{H}$ could be ignored in the strong coupling limit $\eta \rightarrow \infty $, the population approaches $0.5$ asymptotically.


\section{CONCLUDING REMARKS}
\label{sec:conclusion}
This paper developed a new approach to treat the exact open quantum dynamics coupled to harmonic environments.
The approach is derived from HEOM theory and corresponds to a bath collective coordinate mapping approach in the ``hierarchy'' space.
This treatment is more robust than HEOM theory and can treat ultra-strong coupling cases in which the hierarchy size becomes numerous and moves beyond the limits of current computational resources.
We demonstrate our new approach through two examples, which show its pros and cons.

We only treated the case for Drude spectral densities, which causes the Smoluchowski type quantum Fokker--Planck equation.\cite{ikeda2019jctc}
A Drude spectral density can be regarded as an overdamped limit form of a Brownian spectral density, and the momentum degrees of freedom of collective bath coordinates immediately relax to the thermal equilibrium distribution in the limit and do not appear in the numerical description.
To treat more general spectral densities, we must extend the bath collective coordinate space into phase space including momentum degrees of freedom, which requires the Kramers type quantum Fokker--Planck equations.\cite{caldeira1983pa, tanimura1991pra, tanimura2006jpsj, ikeda2017jcp, ikeda2018cp, ikeda2019jctc}
This extension needs different treatments in the mapping transformation Eq.~\eqref{eq:hrchy-transform-mix}, which will be shown in future works.

Here, we solved Eq.~\eqref{eq:partial-h-fpheom} using the traditional finite difference scheme to simplify the calculation structures.
More advanced schemes for Eqs.~\eqref{eq:partial-h-fpheom} and \eqref{eq:full-fpheom}, such as the discrete variable representation \cite{colbert1992jcp} and the time-dependent density matrix renormalization group method \cite{white1992prl, schollwoeck2004rmp, vidal2004prl, daley2004jsmte, white2004prl} could be more efficient.
Reconstructing the hierarchical form using another set of basis functions may have theoretical advantages for cases with strong coupling.
These problems are left for future investigations.


\begin{acknowledgments}
  This work was supported by JSPS KAKENHI Grant No.~JP20K22520.
  A portion of the calculations was performed on the supercomputers at RCCS (Okazaki), RIIT (Kyushu Univ.), and MASAMUNE-IMR (Tohoku Univ., Project No. 202012-SCKXX-0012).
\end{acknowledgments}

\section*{AUTHOR DECLARATIONS}
\subsection*{Conflict of Interest}
The authors declare no conflict of interest.

\section*{DATA AVAILABILITY}
The data that support the findings of this study are available from the corresponding author upon reasonable request.

\appendix
\section{Multi-dimensional Hermite polynomials}
\label{sec:mshf}
The ``basis functions'' introduced in Eq.~\eqref{eq:basis} can be rewritten as
\begin{align}
  \psi _{\bm{n}}(\bm{x})&=\frac{1}{\sqrt {2^{\left|\bm{n}\right|}\pi \mathstrut {\mathstrut \left|\bm{D}\right|}}}\bar{H}_{\bm{n}}(\bm{x})\exp \left(-\bm{x}^{\mathrm{T}}\bm{D}^{-1}\bm{x}\right),
\end{align}
where $\bar{H}_{\bm{n}}(\bm{x})$ is a multi-dimensional polynomial defined as
\begin{align}
  \begin{split}
    \bar{H}_{\bm{n}}(\bm{x})&\equiv (-1)^{\left|\bm{n}\right|}\exp \left(\bm{x}^{\mathrm{T}}\bm{D}^{-1}\bm{x}\right)\\
    &\quad \times \left(\bm{B}^{\mathrm{T}}\bm{D}^{1/2}\bm{\partial }_{x}\right)^{\bm{n}}\exp \left(-\bm{x}^{\mathrm{T}}\bm{D}^{-1}\bm{x}\right).
  \end{split}
\end{align}
The generating function for the set of these polynomials is
\begin{align}
  \sum _{\bm{n}}\bar{H}_{\bm{n}}(\bm{x})\frac{\bm{t}^{\bm{n}}}{\bm{n}!}&=\exp \left[\bm{t}^{\mathrm{T}}\bm{B}^{\mathrm{T}}\left(2\bm{D}^{-1/2}\bm{x}-\bm{B}\bm{t}\right)\right].
\end{align}
This set satisfies the recurrence relations of
\begin{subequations}
  \begin{align}
    x_{j}\bar{H}_{\bm{n}}(\bm{x})&=\frac{1}{2}\sum _{k}\left(\bm{D}^{1/2}\bm{B}^{\mathrm{T}-1}\right)_{jk}\bar{H}_{\bm{n}+\bm{e}_{k}}(\bm{x})\notag\\
    &\quad +\sum _{k}\left(\bm{D}^{1/2}\bm{B}\right)_{jk}\mathstrut n_{k}\bar{H}_{\bm{n-}\bm{e}_{k}}(\bm{x})\\
    \intertext{and}
    \partial _{j}\bar{H}_{\bm{n}}(\bm{x})&=2\sum _{k}\left(\bm{D}^{-1/2}\bm{B}\right)_{jk}n_{k}H_{\bm{n}-\bm{e}_{k}}(\bm{x}).
  \end{align}
\end{subequations}
These polynomials are orthogonal, as shown by
\begin{align}
  &\int \!d\bm{x}\,\bar{H}_{\bm{n}}(\bm{x})\bar{H}_{\bm{n}'}(\bm{x})\exp \left(-\bm{x}^{\mathrm{T}}\bm{D}^{-1}\bm{x}\right)\notag\\
  &=\sqrt {\mathstrut \pi \mathstrut {\mathstrut \left|\bm{D}\right|}}2^{\left|\bm{n}\right|}\bm{n}!\mathstrut {\prod _{k}(\bm{B}^{\mathrm{T}}\bm{B})_{kk}^{n_{k}}}\delta _{\bm{n},\bm{n}'}.
\end{align}
The $\bar{H}_{\bm{n}}(\bm{x})$ is an extension of the \textit{physicist's Hermite polynomials}, which are characterized by
\begin{align}
  H_{n}(x)&\equiv (-1)^{n}\exp \left(x^{2}\right)\partial _{x}^{n}\exp \left(-x^{2}\right),\\
  \sum _{n}H_{n}(x)\frac{t^{n}}{n!}&=\exp [t\left(2xt-t\right)],
\end{align}
\begin{subequations}
  \begin{align}
    xH_{n}(x)&=\frac{1}{2}H_{n+1}(x)+nH_{n-1}(x),\\
    \partial _{x}H_{n}(x)&=2nH_{n-1}(x),
  \end{align}
\end{subequations}
and
\begin{align}
  \int \!dx\,H_{n}(\bm{x})H_{n'}(x)\exp \left(-x^{2}\right)&=\sqrt {\mathstrut \pi }2^{n}n!\delta _{n,n'}.
\end{align}
The polynomials $\bar{H}_{\bm{n}}(\bm{x})$ relate to the ordinary multi-dimensional Hermite polynomials $H_{\bm{n}}^{[\bm{C}]}(\bm{x})$ \cite{berkowitz1970mc, kauderer1993jmp} as
\begin{align}
  \bar{H}_{\bm{n}}(\bm{x})&=H_{\bm{n}}^{[2\bm{B}^{\mathrm{T}}\bm{B}]}\left(\bm{B}^{-1}\bm{D}^{-1/2}\bm{x}\right),
\end{align}
where $H_{\bm{n}}^{[\bm{C}]}(\bm{x})$ is defined as
\begin{align}
  H_{\bm{n}}^{[\bm{C}]}(\bm{x})\equiv (-1)^{\left|\bm{n}\right|}\exp \left(\frac{1}{2}\bm{x}^{\mathrm{T}}\bm{C}\bm{x}\right)\bm{\partial }_{x}^{\bm{n}}\exp \left(-\frac{1}{2}\bm{x}^{\mathrm{T}}\bm{C}\bm{x}\right).
\end{align}

\section{Evaluation of the residues for higher-order poles}
\label{sec:mr}
This section gives an evaluation formula of the residue for higher-order poles in matrix form, which is useful to evaluate the bath correlation functions in Eqs.~\eqref{eq:S} nad \eqref{eq:A}.
We assume that a function $f(z)$ has a pole of $m$th order at $z=C$, i.e.,
\begin{align}
  f(z)&=\frac{\tilde{f}(z)}{(z-c)^{m}},
\end{align}
where $\tilde{f}(z)$ is a holomorphic function in the neighborhood of $c$.
The residue of $f$ around $z=c$ is generally evaluated as
\begin{align}
  \mathrm{Res}_{f}(c)&\equiv \frac{1}{(m-1)!}\lim _{z\rightarrow c}\frac{d^{m-1}}{dz^{m-1}}\tilde{f}(z).
\end{align}
We introduce an $m{\times }m$ matrix $\tilde{\bm{C}}$ defined as
\begin{subequations}
  \begin{align}
    \tilde{\bm{C}}&\equiv
    c
    \begin{pmatrix}
      1 & 0 & \dots & & 0\\
      -1 & 1 & 0 & \ddots & \\
      0 & \ddots & \ddots & 0 & \vdots \\
      \vdots & \ddots & \ddots & 1 & 0\\
      0 & \dots & 0 & -1 & 1\\
    \end{pmatrix},
  \end{align}
  i.e.,
  \begin{align}
    \left(\tilde{\bm{C}}\right)_{jk}&=c\left(\delta _{j,k}-\delta _{j,k+1}\right).&(1\leq j,k\leq m)
  \end{align} 
\end{subequations}
This matrix satisfies
\begin{align}
  \left[\left(\tilde{\bm{C}}-c\right)^{r}\right]_{jk}&=(-c)^{r}\delta _{j,k+r}.
\end{align}
Therefore,
\begin{align}
  \bm{e}_{m}^{\mathrm{T}}\left(\tilde{\bm{C}}-c\right)^{r}\bm{e}_{1}&=(-c)^{m-1}\delta _{m-1,r}.
\end{align}
Using the above relations gives
\begin{align}
  \begin{split}
    \mathrm{Res}_{f}(c)&=\frac{1}{(m-1)!}\lim _{z\rightarrow c}\frac{d^{m-1}}{dz^{m-1}}\tilde{f}(z)\sum _{r=0}^{m-1}\delta _{m-1,r}\\
    &=\frac{1}{(-c)^{m-1}}\lim _{z\rightarrow c}\bm{e}_{m}^{\mathrm{T}}\sum _{r=0}^{m-1}\frac{1}{r!}\frac{d^{r}}{dz^{r}}\tilde{f}(z)\left(\tilde{\bm{C}}-c\right)^{r}\bm{e}_{1}\\
    &=\frac{1}{(-c)^{m-1}}\bm{e}_{m}^{\mathrm{T}}\tilde{f}(\tilde{\bm{C}})\bm{e}_{1}.
  \end{split}
\end{align}
Here, we used the Buchheim formula,\cite{horn1991book} which gives the value of a function for a general matrix.
In the case of $\tilde{\bm{C}}$ whose eigenvalue is $c$ with multiplicity $m$, the formula for a function $g(z)$ reduces to
\begin{align}
  g(\tilde{\bm{C}})&=\lim _{z\rightarrow c}\sum _{r=0}^{m-1}\frac{1}{r!}\frac{d^{r}}{dz^{r}}g(z)(\tilde{\bm{C}}-c)^{r}.
\end{align}

\section{Projection to system-bath collective coordinate space}
\label{sec:shi}
A joint distribution between the system and bath collective coordinate can be evaluated as
\begin{align}
  \hat{f}(x^{\mathrm{D}},t)&\equiv \hat{f}_{\bm{0}}(x^{\mathrm{D}},t)=\int \!d\bm{x}^{\mathrm{BE}}\,\mathcal{S}^{\mathrm{ren}}\mathcal{S}^{\mathrm{mix}}\tilde{f}(x^{\mathrm{D}},\bm{x}^{\mathrm{BE}},t),
\end{align}
By substituting Eq.~\eqref{eq:expansion}, we have
\begin{align}
  \tilde{f}(x^{\mathrm{D}},\bm{x}^{\mathrm{BE}},t)&=\sum _{n^{\mathrm{D}},\bm{n}^{\mathrm{BE}}}\hat{\rho }_{n^{\mathrm{D}},\bm{n}^{\mathrm{BE}}}(t)\psi _{n^{\mathrm{D}}}(x^{\mathrm{D}})\psi _{\bm{n}^{\mathrm{BE}}}(\bm{x}^{\mathrm{BE}}).
\end{align}
This gives
\begin{align}
  \hat{f}(x^{\mathrm{D}},t)&=\sum _{n^{\mathrm{D}},\bm{n}^{\mathrm{BE}}}\prod _{\xi }{\bm{\sigma }^{(\xi )}}^{\bm{n}^{(\xi )}}\hat{\rho }_{n^{\mathrm{D}},\bm{n}^{\mathrm{BE}}}(t)\Psi _{n^{\mathrm{D}}+\left|\bm{n}^{\mathrm{BE}}\right|}(x^{\mathrm{D}})
  \label{eq:},
\end{align}
with
\begin{align}
  \begin{split}
    \Psi _{n}(x^{\mathrm{D}})&\equiv \frac{1}{\sqrt {\mathstrut \pi }(-g^{f})^{n}\left(\sqrt {2D^{f\mathrm{BE}}_{[\gamma ^{\mathrm{D}}]}}\right)^{n+1}\mathstrut n!}\\
    &\quad \times H_{n}\left(\mathstrut \frac{x}{\sqrt {\mathstrut 2D^{f\mathrm{BE}}_{[\gamma ^{\mathrm{D}}]}}}\right)\exp \left(-\frac{1}{2D^{f\mathrm{BE}}_{[\gamma ^{\mathrm{D}}]}}x^{2}\right),
  \end{split}
\end{align}
where $D^{f\mathrm{BE}}_{[\gamma ^{\mathrm{D}}]}$ is given in Eq.~\eqref{eq:D-variance}.
This is an extension of the formula given in Ref.~\onlinecite{liu2014jcp} except for the scale factor $g^{f}$ introduced to non-dimensionalize $q=x^{\mathrm{D}}$.
If higher-order poles do not exist, i.e., $m^{(\xi )}=1$, the above result reduces to those in Refs.~\onlinecite{liu2014jcp, mangaud2019jcp}.
In the above derivation, we used the formula

\begin{align}
  \begin{split}
    &\exp \left(a\partial _{x}^{2}\right)H_{n}(\sqrt {\mathstrut b}x)e^{-bx^{2}}\\
    &=\mathstrut \frac{1}{(\sqrt {\mathstrut 1+4ab})^{n+1}}\exp \left(-\frac{b}{1+4ab}x^{2}\right)H_{n}\left(\sqrt {\mathstrut \frac{b}{1+4ab}}x\right).
  \end{split}
\end{align}

\let\emph=\textit

\bibliography{ikeda2022}

\end{document}